# NEW HORIZONS SOLAR WIND AROUND PLUTO (SWAP) OBSERVATIONS OF THE SOLAR WIND FROM 11-33 AU

H. A. Elliott[1], D. J. McComas[1,2], P. Valek[1,2], G. Nicolaou[3], S. Weidner[1], and G. Livadiotis[1],

[1]Southwest Research Institute, 6220 Culebra Road, San Antonio, TX 78238, helliott@swri.edu

[2]Department of Physics and Astronomy, University of Texas at San Antonio, San Antonio, TX 78249, USA

[3]Swedish Institute of Space Physics, Box 812, SE-98128, Kiruna, Sweden

SHORT TITLE: NEW HORIZONS SOLAR WIND OBSERVATIONS

ABSTRACT

The Solar Wind Around Pluto (SWAP) instrument on NASA's New Horizon Pluto mission has collected solar wind observations en route from Earth to Pluto, and these observations continue beyond Pluto. Few missions have explored the solar wind in the outer heliosphere making this dataset a critical addition to the field. We created a forward model of SWAP count rates, which includes a comprehensive instrument response function based on laboratory and flight calibrations. By fitting the count rates with this model, the proton density (n), speed (V), and temperature (T) parameters are determined. Comparisons between SWAP parameters and both propagated 1 AU observations and prior Voyager 2 observations indicate consistency in both the range and mean wind values. These comparisons as well as our additional findings confirm that small and midsized solar wind structures are worn down with increasing distance due to dynamic interaction of parcels of wind with different speed. For instance, the T-V relationship steepens, as the range in V is limited more than the range in T with distance. At times the T-V correlation clearly breaks down beyond 20 AU, which may indicate wind currently expanding and cooling may have an elevated T reflecting prior heating and compression in the inner heliosphere. The power of wind parameters at shorter periodicities decreases with distance as the longer periodicities strengthen. The solar rotation periodicity is present in temperature beyond 20 AU indicating the observed parcel temperature may reflect not only current heating or cooling, but also heating occurring closer to the Sun.

Key words: --- solar wind --- shock waves --- interplanetary medium --- corotating interaction regions --- (sun:) solar wind --- sun: rotation --- sun: heliosphere

1. INTRODUCTION

New Horizons (NH) is a planetary Pluto flyby mission, which has also collected solar wind observations with the Solar Wind Around Pluto (SWAP) instrument. Few missions have journeyed to the outer heliosphere and only Voyager 2 and Pioneers 10 & 11 missions have extensive solar wind coverage beyond 10 AU (Figure 1). Therefore, these SWAP solar wind observations are a critical addition to the field. The SWAP instrument began extensive solar wind coverage beyond ~22 AU, and these observations continue to be collected with excellent coverage beyond the Pluto flyby at 32.9 AU. We describe the SWAP instrument, calibration, measurements, and the analysis performed to determine the solar wind proton density (n), speed (V), and temperature (T). This paper documents this dataset, provides a variety of tests to insure the data quality, and examines the solar wind properties over the little sampled range from 22-33 AU. We compare SWAP wind parameters to propagated measurements collected at 1 AU, and to prior Voyager 2 measurements. In addition to verifying our analysis, these comparisons reveal that large-scale features on the order of the solar rotation period or greater observed at 1 AU are identifiable beyond 20 AU; however, smaller-scale structures are significantly worn down. Evidence of the dynamic interaction of differing-speed wind parcels wearing down small-scale structures with increasing distance from the Sun is also apparent in our analysis of both the periodicities in the solar wind parameters and the proton temperature and speed relationship.

2. INSTRUMENTATION AND COVERAGE

The SWAP instrument (McComas et al. 2008) is a top-hat electrostatic analyzer (ESA) with two Channel Electron Multiplier (CEM) detectors. Ions enter the aperture, pass through a Retarding Potential Analyzer (RPA) then the ions within a given energy range are bent through the ESA. Next, the ions encounter a foil where any electrons scattering from the front foil surface are subsequently detected by the secondary CEM, and the ion and any electrons scattering from the back of the foil are detected by the primary CEM. The primary and secondary measurements combined produce very low background coincidence detections. The voltages for the ESA, which control the energy passband are swept in order to obtain an energy per charge count rate distribution. First a coarse scan is performed using large steps such that a distribution is produced for the full instrument energy range (~21-7800 eV). Then a fine scan is performed using smaller voltage steps centered on the step where the peak count rates were observed in the prior coarse scan. Both the coarse and fine scans have 64 steps and the count rates are accumulated for 0.39 seconds at each step. A detailed description of the instrument is given by McComas et al., (2008).

The SWAP solar wind coverage before 2012 was limited to about 1.5 months prior to the approach of Jupiter, and short (1 day to 4 weeks) annual spacecraft checkout intervals because the original plan was for all instruments to be off for most of the cruise phase to Pluto. In late 2011 with National Aeronautics and Space Administration (NASA) approval, the final operational plans and flight tests were successfully completed to allow SWAP to continue to operate during previously scheduled "hibernation" intervals. This paper focuses on the solar wind measurements made after the Jupiter flyby (after 2008 October 7) when NH was far enough away from the Sun that it was not necessary to use the RPA to reduce the energy passband to protect the instrument from solar wind fluxes too high for an instrument designed to measure the much lower solar wind flux near Pluto (McComas et al. 2007). Therefore, we use the observations when the RPA was off and only the ESA was on. With the enhanced solar wind coverage beginning in early 2012 at ~22 AU, we are able to perform more complete statistical analysis of these measurements.

## 3. TRAJECTORY AND INSTRUMEN FIELD-OF-VIEW

Coincidently, NH is moving along nearly the same longitude as Voyager 2 (Figure 1) and stays at low heliographic inertial latitudes (|HGI Lat.| < 7˚). When Voyager 2 was collecting observations over the distance range of these SWAP observations (11-33 AU), it was also at low latitudes (|HGI Lat.| < 4˚). By 2012, the spacecraft speed had reached an asymptote of about 14 km s$^{-1}$ and continues at nearly that speed (Figure 2). The SWAP instrument is positioned on the New Horizons spacecraft such that the center of the field-of-view (FOV) is closely aligned with the spacecraft antenna, which is also along the spacecraft +Y axis (Figure 3). The SWAP field of view is approximately 10˚ by 276˚ based on the full width at half maximum of the instrument response. The large angular dimension is in the X-Y plane. The normal to the top of the instrument is well aligned with the –Z spacecraft axis. The large angular dimension is given the symbol ϕ (azimuth angle), and the narrow angular dimension is given the symbol θ (polar angle). During cruise, the spacecraft spins about this +Y axis and points toward Earth. Beyond 11 AU (20 AU) the Sun-probe-Earth angle is less than 6˚(3˚); therefore, for most of the spinning observations presented here the Sun is in the FOV since most of the observations are beyond 20 AU. During the spinning intervals, SWAP has excellent coverage of the sunward facing direction providing continuous solar wind observations during cruise and other spinning intervals.

Figure 4 shows the observations since the beginning of 2012 when the increased coverage began. Each panel is an energy-time spectrogram with the coincidence count rates color-coded for a given year. These observations are coarse energy sweeps consisting of 64 steps and the count rates at each step are accumulated for 0.39 seconds. On the bottom panel the data extends to 15:49 UT on 2015 August 25. There are several noteworthy aspects of this figure. Both the proton (typically red) and alpha (green-blue) beams are clearly observed along with the interstellar pickup protons (typically a medium shade of blue) (McComas et al. 2010; Randol et al. 2012; 2013). At these distances, the solar wind is typically cold enough that there is a clear separation between the proton ($H^+$) and alpha ($He^{++}$) peaks allowing the interstellar pickup protons to be observed between the beams in addition to being observed above the alpha peak. Being able to acquire these interstellar pickup observations given the accumulation time for each energy step is only 0.39 seconds reflects the large SWAP field of view and high sensitivity. Many shocks are observed such as the one on 2012 July 27. Another dominant feature are long quiet intervals such as in early 2013 where the nature of structures in the solar wind seems worn down compared to what is typically observed in the inner heliosphere.

Although the coarse scans shown in Figure 4 seem straightforward, more complexity in the count rate distributions are revealed in the fine energy sweeps where the energy step size is significantly smaller. Figure 5 shows a coarse-fine sweep pair taken at about 11.47 AU when the spacecraft was spinning. Small fluctuations in the count rate seem to occur at specific angles in $\theta$ and $\phi$. In the next section the instrument response is described and incorporated into a forward model of the count rates. Once the angular instrument response functions are included, these fluctuations in the count rates are accounted for since the count rates are then consistently well simulated over a wide variety of pointing configurations.

## 4. COMPREHENSIVE FORWARD MODEL

The basic approach by which the SWAP solar wind parameters are produced is through the use of a forward model. Instrument response functions are combined with a Maxwellian distribution to produce a model of the detected count rates. The n, V, and T are adjusted to find a minimum in the $\chi^2$ between the model and data. Equation 1 provides the integration required to model the count rate (C) for a given energy step ($E_{step}$), and the n, T, and V fit parameters are contained in the distribution function (f).

$$C(E_{step}) = \int_{\phi_c-\Delta\phi/2}^{\phi_c+\Delta\phi/2} \int_{\theta_c-\Delta\theta/2}^{\theta_c+\Delta\theta/2} \int_{v_c-\Delta v/2}^{v_c+\Delta v/2} Aw(v,\theta)p(\phi)f(v,\theta,\phi)(\cos\theta)v^3 dv d\theta d\phi \quad (1)$$

For the solar wind analysis presented here, we use a drifting Maxwellian for $f(v,\theta,\phi)$, and as an initial guess the bulk velocity is assumed to be in the radial direction from the Sun. In Equation 1 the $\theta$ and $\phi$ angles are the same polar and azimuthal angles defined in Figure 3, and the angle integrations are over the SWAP field of view where $\theta_c$ and $\phi_c$ are the instrument boresight. The speed integration is over the energy range for a single step spanning $\Delta v$ centered on the center speed ($v_c$). The integrations are performed numerically using 5 segments for a given speed step, 20 segments for the 10° polar angle width ($\Delta\theta$), and 14 segments in the azimuthal direction. Since the solar wind beam is relatively cold, we do not need to integrate over the full azimuthal width ($\Delta\phi$), we only integrate within +/-15 ° of where the wind beam is centered. The other quantities in count rate expression (Equation 1) represent the instrument response. The instrument response consists of the effective area (A), detector efficiency ($E_{detector}$), the energy-polar response (w(v,$\theta$)), and azimuthal response (p($\phi$)).

### 4.1. INSTRUMENT RESPONSE

The instrument response functions are derived from analysis of both pre-flight laboratory and inflight calibration data. The overall sensitivity of the instrument is related to the product of a fixed area (A) determined to be 0.374 cm$^2$ in the laboratory, and a time variable detection efficiency ($E_{detector}$(t)) determined periodically inflight. Since the detection efficiency changes with CEM usage over time, we determine the coincidence detection efficiency by performing regular inflight gain tests, which are usually performed a couple times a year and are ~1.5 hours in duration. During the gain tests the detector voltages are adjusted, and the observed count rates are recorded. As the CEM voltage is increased the count rates typically rise until a plateau is reached. The operational voltage for science operations is set below this plateau. The efficiency is typically determined as the ratio of the count rate at a given CEM voltage to the level of the plateau. We compared this traditional method to a newer method by Funsten et al., (2005) where the coincidence detection efficiency is defined as the coincidence rate ($N_C$) squared divided by the primary ($N_P$) and secondary ($N_S$) rate such that $E_{detector} = N_c^2/(N_S N_P)$. Note that this method does not have any mass dependence. The Funsten method for determining the efficiency provides an absolute calibration, which is a significant improvement over many electrostatic analyzer measurements that do not have a means of absolute calibration. In Figure 6 we show the Funsten method results during the gain tests. Although this method could be applied for every direct measurement, currently we are only using the values derived during the gain tests. We created a lookup table of the coincidence efficiency as a function of time for the CEM operational voltages being used to collect the science observations (shown as the thick line in Figure 6).

The speed-polar response ($w(v,\theta)$) is based on a combination of lab calibration and electro-optic ion trajectory simulation (Figure 7). This response can be expressed in terms of energy instead of speed. To determine this response function the instrument was mounted in a laboratory calibration chamber and rotated in the polar direction for a given calibration beam energy with the azimuthal angle fixed at 0°. At a given polar angle, ESA energy sweeps were preformed with SWAP while measuring the count rates. The detected rates vary significantly in both θ and energy. An example of the response at 1 keV is shown on the left panel of Figure 7 where the y-axis is the ratio of the beam energy ($E_{beam}$) and the energy at a given energy step ($E_{step}$). Lab calibrations are necessarily limited, so we use them to validate results from an electro-optical trajectory simulation (SIMION®) software and then use these simulations to characterize the SWAP instrument response over energy and θ (Nicolaou et al. 2014). The right panel of Figure 7 shows the simulated response, which clearly matches the measured response. All the simulations for different energy steps were performed with an equal number of initial ions. The final set of simulations were scaled such that the simulation at 1 keV matched both the shape of the transmission as a function of θ, and the simulated y-intercept of the peak of the energy-polar curve matched the laboratory data. From the simulations, we created a large lookup table of the energy-polar response arrays for each energy step used in the forward comprehensive count rate models.

The azimuthal response is based on the final full FOV response calibration, which was performed prior to the final delivery of the SWAP instrument. This occurred after the CEM detectors were refurbished with the flight CEMs, which had completed an extensive initialization process (McComas et al. 1987). This calibration was performed with a beam energy of ~1 keV where the beam was periodically measured with a separate beam monitor. The instrument was rotated using a pair of mounting stages and the stage motion information was used to determine when the beam was at each θ and ϕ location. The FOV response measurements had a 1° resolution in the polar dimension and 1°-2° in the azimuthal dimension. The 1° azimuthal resolution tests were binned down to 2° in order to combine several tests and provided complete coverage in θ and ϕ. The final FOV response array was normalized by the rate at 0° in both θ and ϕ since other calibration factors were determined at this location (Figure 8). To obtain a final azimuthal response, the center bins in θ from -1.5 to 1.5° were averaged and normalized to the value at 0° in ϕ (Figure 9). This azimuthal response curve in Figure 9 is used in the comprehensive forward count rate model.

### 4.2. VIEWING DIRECTION

Many of the expressions in the comprehensive model depend on the solar wind beam entry angles. When the spacecraft is spinning, the Sun's location essentially makes a circle on a θ and ϕ plot (e.g. Figure 10 left panel). In the right panel of Figure 10 we show the location of the Sun determined from the spacecraft attitude data for θ (black) and ϕ (red) as a function of time since the start of a sweep. The solid lines connecting the points are from a fit using the parametric form of the equation for a circle. Clearly, this fit reproduces the Sun location. To insure that the location of the Sun can be accurately described as a circle when the spacecraft is spinning, we also fit all the attitude measurements using the parametric equations for an ellipse (e.g. Foerster 2003; 2003). We made a histogram of the ratio of the semi-major to semi-minor axes (amplitude in the polar and azimuthal direction). Indeed, we observe a tight distribution about a value of 1 indicating that fitting using the parametric circle equations is sufficiently accurate (Figure 11). We use parametric circular fits to the Sun location to find the origin of the circle. The individual origin values for the individual sweeps (black) shown in Figure 12 are averaged over a 3-day window (red) to create the lookup table. Since the origin is predetermined, it can be held fixed while fitting the count rates to find the density speed, and temperature.

For the spinning measurements when attitude data is available the initial guess for the location of solar wind is based on the circle the Sun makes in the SWAP FOV. The size of the circle and the phase of the wind are then additional fit parameters. When no attitude data is available while spinning, the solar wind again is assumed to make a circle in the SWAP FOV. The initial guess for the radius is based on the Sun-Probe-Earth angle with the radius and phase as fit parameters. The origin for of the circle is held fixed using values from the predetermined table (Figure 12). In contrast to the spinning times, when the spacecraft is in 3-axis stabilized pointing, the circle approximation is no longer valid. For the 3-axis stabilized times the initial guess for the solar wind direction is for the Sun location in $\theta$ and $\phi$, and additional fit parameters $\delta\theta$ and $\delta\phi$ define the offset of the solar wind from the Sun direction.

### 4.3. FITTING PROCEDURE

The initial guesses for the comprehensive model are based on results from fitting with a simplified analytic model described in Appendix A. Figure 13 shows an overview of the overall fitting procedure. For both models the basic fitting is performed with the MPFIT IDL routines by Markwardt (2009) based on the Levenberg-Marquardt algorithm (e.g. Moré 1978; 1978). The analytic model does not include the w(v,θ), and p(φ) responses, and consequently does not reproduce the fluctuations that correspond to angle at which the solar wind ions enter the instrument. Therefore, the derivation for the analytic model starts with Equation 2.

$$C(E_{step}) = \int_{\phi_c-\Delta\phi/2}^{\phi_c+\Delta\phi/2} \int_{\theta_c-\Delta\theta/2}^{\theta_c+\Delta\theta/2} \int_{v_c-\Delta v/2}^{v_c+\Delta v/2} Af(v,\theta,\phi)(\cos\theta)v^3 dv d\theta d\phi \quad (2)$$

Then the integrations are simplified to obtain Equation 3 as the final form of the analytic model

$$C(E_{step}) = \left(nA\left(\frac{\beta}{\pi}\right)^{\frac{3}{2}} e^{-\beta(v_c^2+u^2-2uv_c)}\right)\left(\sqrt{\frac{\pi}{\beta uv_c \sin\theta_c'}} erf\left(\sqrt{\beta uv_c}\left(\frac{\Delta\phi}{2}\right)\right)\right)\left(\sin^{-1}\left(\frac{v_{th}}{v_c}\right)v_c^4 \frac{\Delta v}{v_c}\right) \quad (3)$$

where $\boldsymbol{\beta = m/2kT}$, m is the mass, and k is the Boltzmann constant.

Since the analytic model is fast, we run the analytic model multiple times to explore a wide range of possible n, V, and T values in order to have an initial starting point for the comprehensive model with a lower $\chi^2$ value. For each run we use the speed corresponding to the peak count rate as the initial speed guess; however, we use 40 different pairs of temperature (100 to $3\times10^6$ K) and density ($1\times10^4$ to 0.65 cm$^{-3}$) as initial guesses which span the range of possible values. Each of these fits produces a set of fit parameters optimized to minimize the $\chi^2$ between the model results and actual SWAP data, and we choose the set with the lowest $\chi^2$. Several other optimizing steps for the analytic procedure are outlined in Figure 14. The last step of the analytic model procedure is to scale analytic fit results for n and T using scale factors based on comparing analytic and comprehensive results (Appendix A).

The comprehensive fit procedure is diagramed in Figure 15. We begin by comparing the simulated count rates using comprehensive model with initial guesses from the analytic process to rates using the prior time step values as initial guesses. The set of fit parameters producing the lowest $\chi^2$ with the data is selected. The comprehensive fit procedure has slightly different optimizing procedures for spinning and 3-axis stabilized intervals that are detailed in Figure 15. The final comprehensive data set is filtered to ensure that the final results are based on fits that accurately reproduce the count rate distributions. We remove any data points when Sun is outside the field of view, reduced chisquare ($\boldsymbol{\chi_r^2}$) is greater than 100, or either the Pearson or Spearman correlation coefficients are < 0.93.

Some example fit results are shown in Figure 16 at several different distance ranges from the Sun when the spacecraft was spinning. The analytic model (red), which does not account for the angular response does, not reproduce the small fluctuations in the proton peak count rates (black and purple). However, the comprehensive model (blue) does a good job of reproducing these such the $\boldsymbol{\chi_r^2}$ is low and the correlation coefficient is high. This is also the case during hibernation intervals when no attitude data is available. We show such a hibernation example in the third column of Figure 16. During the cruise intervals, no attitude data is collected since those intervals were originally intended to be hibernation intervals. For intervals when no attitude data is available as in the third column of Figure 16, it is still possible to obtain good fits by estimating the radius of the circle using the Sun-probe-Earth angle.

To assess how well the fitting procedure works, we explore the parameter space around the final fit parameters. We find that when the speed is varied the $\chi_r^2$ steeply rises as the speed is adjusted either above or below the final value (Figure 17). If we adjust both the density and temperature values above and below the final values, the $\chi_r^2$ does not rise as sharply as it does for the speed. In order to explore how the fit parameters are interrelated, we vary two of the fit parameters at the same time and calculate the $\chi_r^2$. We obtain the same basic result as when adjusting one parameter at a time. The $\chi_r^2$ values sharply vary with changes in the speed and less sharply with changes in density and temperature. There is some correlation between the density and temperature parameters; however, it is the shallowness of the $\chi^2$ minimum, which makes finding the density and temperature difficult since the Levenberg-Marquardt algorithm searches the parameter space using the gradients $\chi^2$ for a given parameter (Moré 1978). It is much easier to find the speed given that the solar wind is a cold beam, the peak of the count rate distribution is well defined as a function of energy per charge, and the speed corresponding to the peak location is close to the bulk speed. We alleviate the shallow minimum problem for density and temperature by first performing analytic fits with initial guesses spanning the reasonable range of possible densities ($1\times10^4$ to 0.65) and temperatures (100 to $3\times10^6$). Since the peak speed is so close to the final speed found, and minimum for $\chi_r^2$ is so narrow and deep, we do not need to search over a wide range of speeds.

For a simple 1-D Maxwellian distribution as a function of the speed, the bulk speed is essentially at the location of the peak in the distribution, the width of the distribution is related to the temperature, and the density is related to the area under the distribution. SWAP measures the count rate of ions and this is directly related to the flux of ions that enter the aperture. For a given energy step the SWAP instrument allows ions with a given range in energy (ΔE) to enter anywhere inside the large 10° by 276° aperture. These ions are bent through the ESA dome and focused onto the foil, and the CEM detectors are on either side of this foil. Since ions can enter over the full FOV and are focused to a small detection area, the SWAP instrument observes over a large swath of the sky. This means there is not quite a one to one relationship between the count rate distribution and the solar wind parameters as there is for an ideal Maxwellian distribution function. However, to a zeroth order those relationships do hold. In the top row of Figure 18 we show energy-time coincidence count rate distribution plot, and corresponding final solar wind speed, density, and temperature time series below. The two plots represent two contrasting intervals. On the left side of Figure 18 is an active interval with clear shocks, compressions, and rarefactions. On the right side is a quiet interval with nearly constant solar wind speed, but distinct changes in the width of the distribution and count rate. One can see that the derived speed tracks the location of the proton peak, the temperature tracks the width of the distribution, and the density tracks the overall coincidence rate of the proton peak.

In the active interval there is a shock on 2012 July 27 (209) and a clear drop in density spanning from 2012 October 10 to 2012 October 16 (284-290). In the quiet interval even though the solar wind speed is steady, enhancements in the distribution width are coincident with enhancements in temperature (e.g. the enhancement centered on 2013, March 2 (061)). In the inner heliosphere the solar wind temperature and speed are strongly positively correlated (Neugebauer & Snyder 1966; Elliott et al. 2012). The solar wind speed being more steady and the temperature being elevated may reflect that dynamic interaction between fast wind parcels running into slower wind parcels emitted earlier in time in the inner heliosphere causes the fast wind to slow and the slow wind to speed up while simultaneously heating the wind. This would create some solar wind intervals beyond 25 AU where the solar wind speed is steady and the temperature is elevated. If the solar wind structures had only modest speed differences closer to the Sun then the structures could easily be worn down by 25 AU.

## 5. RADIAL TRENDS IN SOLAR WIND PROPERTIES

Examining the radial trends in the SWAP observations is challenging since these observations begin around 11 AU, and are very sparse between 11 and 22 AU. The solar wind density and temperature are known to decrease rapidly with distance in the inner heliosphere; therefore, not having inner heliospheric measurements from SWAP is a significant disadvantage. To help compensate for this, we perform power law fits using only observations after 2012 January 28 (22.21-33.26AU) when the first long cruise observations began, and hold the amplitude fixed to the average 1 AU values from 2012-2013 determined using ACE and STEREO observations (Figure 19). In the distance range of the SWAP measurements, the most comprehensive solar wind speed measurements to which comparisons can be made are those of Voyager 2. Table 1 summarizes these results and those of from prior Ulysses and Voyager 2 studies. The Ulysses observations only extend to 5.4 AU, but we include them since there are so few radial trend observations in the outer heliosphere. Our temperature exponent is closest to the Voyager 2 fit from 1 to 30 AU, and our density exponent has a lower absolute value than the other density exponents.

Even though the overall exponents are different than those of Voyager 2, a direct comparison SWAP and Voyager 2 measurements appear consistent with one another. In Figure 20 we show the speed, density, temperature and spacecraft latitude as a function of distance from the Sun for both the SWAP (blue) and Voyager 2 (black) solar wind parameters. Three key factors may explain why directly comparing both data sets shows more consistency than one would expect based on the differences in fit results for the power law exponents. The primary reason is likely that each data set has a different percentage of fast and slow wind. Since the average density and temperature are different for fast and slow wind, the results may have been skewed. The second reason is the lack of inner heliospheric observations for NH, where n and T drop sharply with distance. Yet another reason is that the solar wind and the Sun vary as a function of time and both data sets were collected over a long time range in different solar cycles.

### 5.1. COMPARISON TO PROPAGATED 1 AU SPEEDS

In Figure 18 we showed that in late 2012 there was a lot of activity in the solar wind measured at NH, but in early 2013 there were few solar wind structures and the solar wind speed was much more steady. This sharp contrast makes these two intervals ideal for comparing with 1 AU measurements since this contrast should also be present at 1 AU. Another advantage of this timeframe is that ACE and both STEREO spacecraft, which are all at 1 AU and in the ecliptic plane, were well separated in longitude. We calculate running solar rotation (25.38 day) averages of the speeds from the 1-hour measurements at 1 AU, and use those to determine the propagation time for the solar wind measured at 1 AU to reach the radial distance of New Horizons. Then we select times when the 1 AU spacecraft were within 70° of the longitude with NH and overplot those segments as shown in Figure 21. The New Horizons measurements are shown on the same plot in orange at the cadence that they were recorded without any kind of averaging. The amount of smoothing that the running rotation averages for the 1 AU measurements provides is coincidentally similar to the amount of wearing down in the solar wind structures that occurred by the time those structures reached New Horizons. Averaging over a solar rotation is logical for comparing measurements collected over the same timeframe since over a given solar rotation, there should be about the same mix of fast and slow wind even if the longitude of observation is somewhat different. This simplistic method of determining the propagation time for each individual data point has limitations. It cannot account for all the nonlinear dynamic interactions that occur en route since one measurement can pass another; however, using the running average lessens this effect compared to if we had used the instantaneous speed measured at 1 AU to calculate the propagation time. Although this method seems to proved reasonable agreement for the speed, it does not account for changes in density and temperature that would occur when parcels interact with one another, and either compress or expand.

## 5.2. IDENTIFYING DYNAMIC INTERACTION SIGNATURES

In addition to there being times the temperature is elevated when the solar wind speed is steady, there are also times when the temperature increases as the solar wind speed decreases. This anti-correlation is opposite that of the normal positive correlation present in the inner heliosphere (Neugebauer & Snyder 1966; Elliott et al. 2012). In Figure 22 we show SWAP speeds and temperatures from mid-2012 through late 2013 as recorded and running averages with color-coding. The speed rises are colored (compressions) orange and the speed decreases (rarefactions) are colored blue. The temperature normalized by distance is color-coded by using a running average of the normalized temperature-speed slope. When the temperature-speed slope is positive, the color is pink. When it is negative, the color is green. Note that several of the green times occur when the speed is decreasing (blue). This could mean that several structures have merged prior to reaching New Horizons. Even though the structure is currently in an expanding rarefaction at NH, part of the material in the current rarefaction may have been in a small compression region in the inner heliosphere. A similar conclusion was reached by the analysis of Voyager 2 observations further out by Lazarus et al., (1999). They found that beyond ~40 AU merged interaction regions (MIRs) had only forward shocks and no forward-reverse shock pairs were found. They speculated that the thermal speed (temperature) enhancements occurring after the forward shocks were the remnants of reverse shocks present in the inner heliosphere. Here we find that this process can also be important at relatively small heliocentric distances.

### 5.3. RADIAL VARIATION IN THE TEMPERATURE-SPEED DISTRIBUTION

The solar wind temperature-speed (T-V) relationship significantly evolves with distance. Inside of 0.5 AU there is a sharp transition between the fast and slow wind that becomes less distinct by 1 AU, and by 4 AU a heated component develops at intermediate speeds owing to the interaction of the fast and slow wind (Elliott et al. 2012). We examine the T-V distribution using SWAP data between 21 and 33 AU. In order to compare with inner heliospheric observations where the temperature is scaled by a factor of $r/r_o$ where $r_o$ is 1 AU (top row Figure 23), we also apply this scaling to the SWAP observations (bottom row Figure 23). The range of speeds present has been significantly reduced beyond 21 AU; however, the overall range of temperatures is only slightly reduced compared to what was found between 4 and 5.4 AU. We perform a linear fit to compare to linear fits at 1 AU where the slopes tend to range from 400 to 800 K s km$^{-1}$. The overall distribution appears to span a narrow range of speeds between 300 and 450 km s$^{-1}$, and the range of temperatures remains broad; therefore, the overall slope of the T-V distribution is steep compared to slopes at 1 AU as found in Elliott et al. (2012). The exact scaling factor is not critical to our point about the changes in the distribution since beyond 21 AU as there is no longer a steep trend in either T or V. Adjusting the radial normalization exponent only shifts the distribution up or down and does not affect the shape. Table 2 shows the linear-fit results using both an exponent of 1 to compare with prior studies and the 0.745 exponents found earlier to normalize the temperatures with distance. These slopes are indeed steeper than those at 1 AU although the T-V correlation is not as strong as what is found at 1 AU since the correlation coefficiets are ≤ 0.4. Pioneer 10 analysis by Gazis (1987) found that the range in both thermal speed (temperature) and the bulk speed is less at ~23 AU compared to what was observed from 9-15 AU where as we found the range of speeds to be more limited with increasing distance than that of temperature.

### 6. RADIAL TRENDS IN PERIODICITY

To assess if the solar rotation rate and harmonics of the solar rotation rate commonly found in the solar wind data at 1 AU are observable beyond 20 AU, we compare periodograms created with 1 AU measurements to those created using SWAP measurements from 21-33 AU. Many algorithms for examining periodicities are only well suited for equally spaced measurements. Since the SWAP data were collected at a variety of sampling intervals, we use techniques suited for unequally time spaced measurements. The time spacing is not always uniform even for the 1 AU observations; therefore, we apply the same techniques for all the data sets.

We use three techniques to examine the periodicities: Lomb-Scargle periodograms, discrete Fourier Transform (DFT), and the CLEAN algorithm, which removes spurious features from the DFT. The Lomb-Scargle periodogram technique works well on unevenly sampled data because it weights the data on a per point basis (Press & Rybicki 1989). It differs from standard Fourier analysis in that the Lomb-Scargle normalized periodogram equation is equivalent to performing a least squares fit of the data expressed in sinusoidal form at a given frequency (Press & Rybicki 1989; Lomb 1976; Scargle 1982). The implementation of the Lomb-Scargle technique is given in Chapter 13 Fourier and Spectral Applications by Press et al. (1992). We also use the Discrete Fourier Transform (DFT) described in Roberts et al. (1987) to calculate power spectra. They refer to DFT spectra as being "dirty" because a Fourier transform of unevenly spaced data is a convolution of the transform of "real" sources at specific source frequencies and the Fourier transform of the windowing data sampling function. The way the data is sampled can create false features in the power spectra. We also use the CLEAN algorithm in the Roberts et al. (1987) paper that was first developed by Högbom, (1974). This cleaning algorithm is an iterative technique that subtracts a fraction of the amplitude for the current maximum peak by modeling the source peak as a Gaussian with a fraction of the amplitude of the measured peak. At each iteration, an updated residual spectrum is created and a full model spectrum with additional modeled Gaussians over the desired frequency range is built. The process continues until the residual spectrum only contains noise. The final cleaned periodogram is then convolved with the last residual periodogram to produce a final clean version of the periodogram.

The 1 AU periodograms are based on ACE-SWEPAM 1 hour observations spanning from 1998 January 23 to 2015 July 4. The mean for each wind parameter is removed prior to determining the periodogram. In Figure 24 we show the periodograms of the speed, temperature, and density using the Lomb-Scargle (L-S), DFT, and cleaned DFT techniques. The spectra for the L-S and DFT methods agree quite well. There are clear peaks associated with solar rotation, which would be 27 days accounting for the motion of Earth around the Sun since ACE is at L1. Little power is observed beyond 35 days for the density and temperature. There is a significant amount of power at periods less than 20 days in all three parameters. In the cleaned spectra these key features remain. Lomb-Scargle periodogram analysis of solar wind parameters by Katsavrias et al. (2012) also indicates that 27, 13.5 and 9 day periodicities are present at 1 AU in n, v, and T when many years of data are analyzed. The wavelet analysis of the same data reveals that the power at these specific periodicities varies significantly throughout a given year and even may not be present for several months (Katsavrias et al. 2012).

We divide the SWAP solar wind parameters into three approximately equal time ranges corresponding to three distance ranges: 21.96-25.80 AU, 25.80-29.54 AU, and 29.54-33.6 AU. For each distance range, we created one figure showing the L-S, DFT, and cleaned DFT periodograms for the speed, temperature, and density. The mean is removed as was done for the 1 AU periodograms, and the radial trends in the temperature and density is removed by normalizing by the radial distance using the exponents found earlier. Over the distance range from 21.96 to 25.80 AU (Figure 25), the power for periods less than 20 days is significantly diminished and enhanced beyond 50 days in the spectra for n, v, and T. When the CLEAN algorithm is applied, many of the peaks between 27 and 50 days do not survive the cleaning process, but there is still a clear peak at the solar rotation rate for all three wind parameters. For the middle distance interval (25.80 to 29.54 AU) the periodograms shown in Figure 26 are more complex, particularly in the L-S and DFT ones. The cleaning procedure eliminates some of the complexity for periods greater than 25 days in all three solar wind parameters. In particular the peak at about 33 days is significantly reduced when cleaned. The speed periodogram shows no clear rotation peak, but there is a rotation peak for the density, and a very strong rotation peak for the temperature. In the farthest distance range from 29.54 to 33.26 AU (Figure 27), the power is quite low for the speed parameter at periods shorter than 20 days. However, a broad lumpy peak spans from 23 to nearly 70 days for the speed in the L-S and DFT periodograms. When the cleaning procedure is performed that broad peak splits into two: one at the sidereal rotation period (25 days) and one at about 43 days. Overall the power seems enhanced at long periods for the speed. The temperature shows very distinct peaks at the rotation period, and at periods slightly less than both half the rotation period (12.5), and a third of the rotation period (8.33). The density appears to have some power at one third of the rotation period and the rotation period in the L-S and DFT periodograms, but the cleaned DFT shows this is diminished and significant power only occurs at periods longer than 50 days.

Overall the temperature seems to retain more of the solar wind rotation rate signature beyond 25 AU than the speed and density parameters. The density seems to retain the least rotation rate signature. Beyond 20 AU the power is enhanced at periods longer than the rotation rate and diminished for periods less than 15 days. Our results on the periodicities are consistent with our observations of the small solar wind structures being worn down with increasing distance from the Sun and with prior outer heliospheric observations from Voyager 2 and Pioneers 10 &11. In analysis of Pioneer data from 10 to 15 AU, Gazis (1987) found the solar wind structure to be less regular compared to observations inside of 9 AU, and they could no longer distinguish forward and reverse shock pairs. In the distance range from 9-30 AU Gazis (1987) found that ~70% the solar wind speed was quite flat, and these flat wind speed intervals often had quasi-periodic enhancements in the density and temperature. The other ~30% of the time they there were significant jumps in solar wind speed where the wind was elevated for 30-120 day stretches. Another study of Pioneer observations by Gazis et al. (1999) showed clear CIR periodic signatures with systematic changes in the wind parameters inside of 8 AU, and the development of more complex merged interaction regions between 8 and 10 AU. However, beyond 36 AU they found irregular enhancements in n and T that are not shock associated, not periodic, and are preceded by only a very small V enhancement. Even farther from the Sun beyond 50 AU they found that the variations in the wind parameters were small with no clear CIR signature, and the main time variations were on the scale of 1-1.3 years (Gazis et al. 1999). Analysis of Voyager 2 data by Burlaga et al. (1997), also indicates changes in the periodicity with distance. Burlaga et al. (1997) found correlated quasi-periodic variations in field strength (B) and n with a 26 day period at 14 AU, but not at 43 AU. They also found the V and T profiles to lack periodicity at 14 AU. However, at 43 AU they found speed jumps with a period of ~26 days and corresponding temperature enhancements which followed the jumps. The periodicities seen in the SWAP data are consistent with these previous Voyager 2 and Pioneer observations.

## 7. SUMMARY AND CONCLUSIONS

We have developed a comprehensive forward model to reproduce the SWAP observations, which we use to determine the best possible density, speed, and temperature values for the solar wind from SWAP. This model includes extensive laboratory and flight calibrations, and the Funsten et al. (2005) method of determining the absolute calibration of the coincidence measurements over time, which improves the accuracy especially of the densities. The comprehensive model can simulate small fluctuations caused by ions entering at different angles, and the instrument angular response function being non-uniform. Additionally, we developed a simplified analytic model, which allows a large fit parameter space to be searched quickly. The fit results from the analytic model provide initial guesses for the comprehensive model. By comparing the analytic and comprehensive results, we developed scale factors that significantly improved the accuracy of the analytic model.

To help verify the comprehensive model, we compare the comprehensive results to propagations of 1 AU observations, and to prior Voyager 2 observations. By overlaying the Voyager 2 and SWAP observations we can clearly see that the measurements span the same range of densities, speeds, and temperatures at the same distance ranges. The comparison between the propagated 1 AU solar wind speeds and the SWAP data reveal that large scale features at 1 AU can be identified beyond 20 AU even though many small scale features merge prior to reaching 20 AU. The high degree of coherency (long autocorrelation times) in the solar wind speed (Elliott et al. 2013) probably plays a roll in enabling the use of a simple running rotation average (~25 days) to accurately reproduce the wearing down of structures from 20-30 AU. The same running rotation average produced too much smoothing when applied to the density and temperature measurements.

Radial evolution of the solar wind owing to dynamic interaction of different wind speed parcels is also evident when we examine the temperature-speed (T-V) distribution at different radial distance ranges. The slope of the T-V distribution steepens as the range of speeds is limited more than the range of temperatures with increasing distance. However, by color-coding the speed-time slope and temperature-speed slope, it becomes apparent that the T-V correlation, which is strong at 1 AU can break down significantly and even be anti-correlated at times beyond 20 AU. For instance we observed intervals beyond 20 AU that had a negative speed-time slope indicative of an expanding rarefaction, but had a positive temperature-speed slope that would have typically been indicative of a compression region in the inner heliosphere. This seems to indicate that a parcel of wind currently expanding and cooling may have an elevated temperature reflecting prior heating and compression in the inner heliosphere. The solar rotation period is less apparent in the solar wind speed and density beyond 20 AU when compared to 1 AU observations, although it still is quite apparent in the temperature. Therefore, the temperature may continue to have a solar rotation periodicity because the current temperature of a given parcel reflects not only whether the wind is currently being heated or cooled, but also if it had been heated closer to the Sun. Thus, the temperature seems to retain more of the history of the wind.

Continued SWAP solar wind observations will provide critical outer heliospheric observations since the Pioneer 10 (Pioneer 11) observations extend to 61 (36) AU with only sparse coverage after 46 (26) AU. The Voyager 2 observations provide excellent coverage beyond the termination shock. However, after 40 AU the Voyager 2 heliographic latitude was greater than 10° South, where as New Horizons is continuing to stay at low latitudes and headed towards the enhanced energetic neutral atom emissions of the ribbon found in Interstellar Boundary Explorer observations (McComas et al. 2009).

ACKNOWLEDGEMENTS:

We are deeply indebted to all of the people who made the Solar Wind Around Pluto (SWAP) instrument and New Horizons mission possible. In particular we thank the computer engineers John Hanley, and Greg Dunn; system engineer M. Tapley; and command sequencers Helen Hart, and Mark Kochte who worked extensively to make the SWAP observations possible. This work was carried out as a part of SWAP Investigation on the New Horizons mission, which is part of NASA's New Frontiers Program and a Heliophysics NASA grant NNX12AB26G.

APPENDIX A: ANALYTIC MODEL

We develop an analytic expression for the Solar Wind Around Pluto solar wind measurements, which we use only to improve the initial solar wind parameter estimates that are subsequently used as initial guesses for the comprehensive forward model. The comprehensive model includes several calibration factors. However, for the analytic expression, we use only instrument effective area, instrument field-of-view (FOV), and energy passband (McComas et al. 2008), and do not include any angular response functions. The detected count rate (C) depends on the incident particle flux (nv) and the instrument response. For the solar wind analysis, we examine the SWAP fine energy sweeps. The fine energy steps overlap in energy; therefore, we need to simulate the count rate for each energy step individually. To model the count rate, a Maxwellian distribution is integrated over the instrument field-of-view (FOV) for a given energy step. Here, we describe the energy steps in terms of the corresponding speed value ($v = \sqrt{2E/m}$). The center speed of a given step is $v_c$, and the speed width is $\Delta v$. Similarly the center of the FOV is at $\theta'$ and $\phi_c$, and the corresponding FOV angular widths are $\Delta\theta$ and $\Delta\phi$. Here we use the colatitude ($\theta'$) rather than the polar ($\theta$) where $\theta' = 90° + \theta$. The range of angles is the same for the polar and colatitude ($\Delta\theta' = \Delta\theta$). The SWAP FOV is narrow in the θ-direction (~10°) and very wide in the φ-direction (~276°). In Equation A1 we show a general expression for the count rate where f(v, $\theta'$, φ) is the particle distribution function and the limits of integration extend over the full instrument FOV and speed (energy) range for a given step.

$$C(E_{step}) = \int_{\theta'_c-\Delta\theta/2}^{\theta'_c+\Delta\theta/2} \int_{v_c-\Delta v/2}^{v_c+\Delta v/2} \int_{\phi_c-\Delta\phi/2}^{\phi_c+\Delta\phi/2} Af(v,\theta',\phi)(\sin\theta')v^3 dv d\theta' d\phi \quad (A1)$$

For our analytic expression, we use a drifting Maxwellian distribution function (Equation A2) where $\beta = m/2kT$.

$$f(v,\theta',\phi) = n\left(\frac{\beta}{\pi}\right)^{3/2} e^{-\beta\left((v_x-u_x)^2+(v_y-u_y)^2+(v_z-u_z)^2\right)} \quad (A2)$$

The Cartesian velocity components are expressed in terms of spherical coordinates (Equation A3),

$$\begin{aligned} v_x &= v\sin\theta'\cos\phi \\ v_y &= v\sin\theta'\sin\phi \quad (A3) \\ v_z &= v\cos\theta' \end{aligned}$$

and we center the solar wind beam on the center of the instrument FOV, which is at $\theta'_c = 90°$ and $\phi_c = 0°$, and along the x-axis (Equation A4).

$$\begin{aligned} u_x &= u \\ u_y &= 0 \\ u_z &= 0 \end{aligned} \quad \text{(A4)}$$

Substituting Equation A4 into the distribution simplifies to

$$f = n \left(\frac{\beta}{\pi}\right)^{3/2} e^{-\beta((v_x-u)^2+v_y^2+v_z^2)}. \quad \text{(A5)}$$

Now, we expand the exponent (Equation A6 and A7) using Equation A3.

$$Q = (v_x - u)^2 + v_y^2 + v_z^2 \quad \text{(A6)}$$
$$Q = v^2 \sin^2\theta' \cos^2\phi + u^2 - 2uv\sin\theta'\cos\phi + v^2\sin^2\theta'\sin^2\phi + v^2\cos^2\theta' \quad \text{(A7)}$$

Using trigonometric identities the exponent can be simplified.

$$Q = v^2 + u^2 - 2uv\sin\theta'\cos\phi \quad \text{(A8)}$$

The distribution function then becomes

$$f = n \left(\frac{\beta}{\pi}\right)^{3/2} e^{-\beta(v^2+u^2-2uv\sin\theta'\cos\phi)}. \quad \text{(A9)}$$

Next, we substitute the distribution function into the rate expression (Equation A1) in order to integrate in spherical coordinates.

$$C(E_{step}) = \int_{\theta'_c-\Delta\theta/2}^{\theta'_c+\Delta\theta/2} \int_{v_c-\Delta v/2}^{v_c+\Delta v/2} \int_{\phi_c-\Delta\phi/2}^{\phi_c+\Delta\phi/2} nA \left(\frac{\beta}{\pi}\right)^{3/2} e^{-\beta(v^2+u^2-2uv\sin\theta'\cos\phi)} \sin\theta' \, v^3 \, d\phi \, dv \, d\theta'$$
(A10)

Before integrating we rearrange the terms in the rate expression to collect the expressions, which have a φ-dependence.

$$C(E_{step}) = \int_{\theta'_c-\Delta\theta/2}^{\theta'_c+\Delta\theta/2} \left( \int_{v_c-\Delta v/2}^{v_c+\Delta v/2} \left( nA \left(\frac{\beta}{\pi}\right)^{3/2} e^{-\beta(v^2+u^2)} \int_{\phi_c-\Delta\phi/2}^{\phi_c+\Delta\phi/2} (e^{\beta 2uv\sin\theta'\cos\phi} d\phi) v^3 dv \right) \sin\theta' d\theta' \right) \quad (A11)$$

We now can evaluate the φ integral shown below by letting $a = 2\beta uv\sin\theta'$.

$$\int_{\phi_c-\Delta\phi/2}^{\phi_c+\Delta\phi/2} e^{\beta 2uv\sin\theta'\cos\phi} d\phi = \int_{\phi_c-\Delta\phi/2}^{\phi_c+\Delta\phi/2} e^{a\cos\phi} d\phi \quad (A12)$$

This integral is rather cumbersome so we approximate cosφ using a simple 2-term series expansion.

$$\cos\phi \cong (1 - \phi^2/2 + \cdots) \quad (A13)$$

Using this expansion the φ integrals can be transformed into one, which is easy to evaluate.

$$\int_{\phi_c-\frac{\Delta\phi}{2}}^{\phi_c+\frac{\Delta\phi}{2}} e^{a\cos\phi} d\phi = \int_{\phi_c-\frac{\Delta\phi}{2}}^{\phi_c+\frac{\Delta\phi}{2}} e^{a\left(1-\frac{\phi^2}{2}\right)} d\phi = e^a \int_{\phi_c-\frac{\Delta\phi}{2}}^{\phi_c+\frac{\Delta\phi}{2}} e^{-\frac{a}{2}\phi^2} d\phi \quad (A14)$$

Next, we set $\phi_c = 0°$ because the solar wind beam is centered on the center of the FOV. Since the φ integrand is both an even function, and the limits of integration are symmetric about zero, the full integral is equal to twice the value integrated over half the range of integration. The φ integral is then expressed in error function form.

$$e^a \int_{-\frac{\Delta\phi}{2}}^{\frac{\Delta\phi}{2}} e^{-\frac{a}{2}\phi^2} d\phi = 2e^a \int_0^{\frac{\Delta\phi}{2}} e^{-\frac{a}{2}\phi^2} d\phi = 2e^a \frac{1}{2}\sqrt{\frac{\pi}{a\backslash 2}} erf\left(\sqrt{\frac{a}{2}}\left(\frac{\Delta\phi}{2}\right)\right) \quad (A15)$$

$$\int_{\phi_c-\frac{\Delta\phi}{2}}^{\phi_c+\frac{\Delta\phi}{2}} e^{\beta 2uv\sin\theta'\cos\phi} d\phi = e^{2\beta uv\sin\theta'} \sqrt{\frac{\pi}{\beta uv\sin\theta'}} erf\left(\sqrt{\beta uv\sin\theta'}\left(\frac{\Delta\phi}{2}\right)\right) \quad (A16)$$

Substituting this result into the rate expression (Equation A11) the rate expression becomes

$$C(E_{step}) = \int_{\theta'_c-\frac{\Delta\theta}{2}}^{\theta'_c+\frac{\Delta\theta}{2}} \left(\int_{v_c-\frac{\Delta v}{2}}^{v_c+\frac{\Delta v}{2}} \left(nA\left(\frac{\beta}{\pi}\right)^{\frac{3}{2}} e^{-\beta(v^2+u^2-2vu\sin\theta')} \sqrt{\frac{\pi}{\beta uv\sin\theta'}} erf\left(\sqrt{\beta uv\sin\theta'}\left(\frac{\Delta\phi}{2}\right)\right) v^3 dv\right) \sin\theta' d\theta'\right). \quad (A17)$$

Evaluating the speed integral using the mean value theorem, which approximates the area as a rectangle, the rate becomes

$$C(E_{step}) = \int_{\theta'_c-\frac{\Delta\theta}{2}}^{\theta'_c+\frac{\Delta\theta}{2}} \left(nA\left(\frac{\beta}{\pi}\right)^{\frac{3}{2}} e^{-\beta(v_c^2+u^2-2v_c u\sin\theta')} \sqrt{\frac{\pi}{\beta uv_c\sin\theta'}} erf\left(\sqrt{\beta uv_c\sin\theta'}\left(\frac{\Delta\phi}{2}\right)\right) v_c^3 \Delta v\right) \sin\theta' d\theta'. \quad (A18)$$

Since the integrand for the θ integral is a peaked function, we approximate the area as a triangle rather than a rectangle.

$$C(E_{step}) = \left(nA\left(\frac{\beta}{\pi}\right)^{\frac{3}{2}} e^{-\beta(v_c^2+u^2-2v_c u\sin\theta'_c)} \sqrt{\frac{\pi}{\beta uv_c\sin\theta'_c}} erf\left(\sqrt{\beta uv_c\sin\theta'_c}\left(\frac{\Delta\phi}{2}\right)\right) v_c^3 \Delta v (\sin\theta'_c) \Delta\theta/2\right) \quad (A19)$$

Rearranging terms we obtain the following analytic expression for the rate for a given energy step:

$$C(E_{step}) = \left(nA\left(\frac{\beta}{\pi}\right)^{\frac{3}{2}} e^{-\beta(v_c^2+u^2-2v_c u\sin\theta'_c)}\right) \sqrt{\frac{\pi}{\beta uv_c\sin\theta'_c}} erf\left(\sqrt{\beta uv_c\sin\theta'_c}\left(\frac{\Delta\phi}{2}\right)\right) \left(v_c^4 \frac{\Delta v}{v_c} \sin\theta'_c \frac{\Delta\theta}{2}\right). \quad (A20)$$

Since $\sin\theta'_c = 1$, Equation A20 becomes

$$C(E_{step}) = \left(nA\left(\frac{\beta}{\pi}\right)^{\frac{3}{2}} e^{-\beta(v_c^2+u^2-2v_c u)}\right) \sqrt{\frac{\pi}{\beta uv_c}} erf\left(\sqrt{\beta uv_c}\left(\frac{\Delta\phi}{2}\right)\right) \left(v_c^4 \frac{\Delta v}{v_c} \frac{\Delta\theta}{2}\right). \quad (A22)$$

Letting $\Delta\theta = 2\sin^{-1}(v_{th}/v_c)$ and $v_{th} = \sqrt{1/\beta}$ the final rate equation becomes

$$C(E_{step}) = \left(nA\left(\frac{\beta}{\pi}\right)^{\frac{3}{2}} e^{-\beta(v_c^2+u^2-2v_c u)}\right)\sqrt{\frac{\pi}{\beta u v_c}}\, erf\left(\sqrt{\beta u v_c}\left(\frac{\Delta\phi}{2}\right)\right)\left(v_c^4\, \frac{\Delta v}{v_c}\sin^{-1}(v_{th}/v_c)\right). \quad (A23)$$

The simplifications and assumptions used to derive the analytic expression are flawed; however, this model is only meant to help limit the parameter search space to provide better initial guesses for the full comprehensive model. Later, we show a comparison between the analytic and comprehensive results. The analytic model is systematically different from the comprehensive model so we have developed scale factors to partially correct the analytic model values. The source of the bulk of the systematic error seems to be related to the simplifications in the integrations. Figure A1 shows count rates simulated using the analytic model (Equation A23) as thin lighter colored lines. A simplification of the comprehensive model with only the effective area portion of the instrument response and none of the angular response functions included (Equation A1) is shown as thick darker lines. Both the analytic model and the simplified comprehensive model assume the solar wind beam enters the center of the field-of-view and both are run with same detector efficiency value. When the solar wind beam temperature is varied and the integrations are done numerically, the count rate distribution appears wider and higher than the analytic model (left Figure A1). At low temperatures the numerical integration produces a peak with a flat top and the analytic model has a sharper peak. When the density is varied (right Figure A1) the distribution shape is the same for the analytic model and the numerical integration model, but again the distribution height and width are larger when the integrations are performed numerically. Since the analytic model consistently produces narrower and lower count rate curves owing to the simplifications in the integrations when using this model to fit the data the density and temperature will be systematically overestimated. To improve the initial guesses for the comprehensive models, we developed relationships to scale the analytic results to better match the comprehensive model using observations from 2008 October 10 to 2014 July 4 (Figure A2). After that point, the analytic model fit results were scaled to improve the initial guesses for the comprehensive model.

FIGURE CAPTIONS:

*Figure 1: The trajectory of New Horizons and other missions that explored the outer heliosphere in Heliocentric Aries Ecliptic J2000 (HAE-J2000) coordinates. This coordinate system is a heliocentric system with the Z-axis normal to the ecliptic plane and the X-axis points toward the first point of Aries on the Vernal Equinox, and the Y-axis completes the right-handed system. The interstellar wind direction is from (McComas et al. 2015).*

*Figure 2: New Horizons spacecraft speed as a function of time in years (black). In red is shown the component of the speed along the direction radial from the Sun.*

*Figure 3: The center of the Solar Wind Around Pluto (SWAP) field-of-view (FOV) is aligned with the spacecraft +Y-axis (left panel). The middle panel illustrates SWAP Cartesian coordinates. The right panel illustrates the corresponding SWAP angles. The θ polar angle is positive towards +Z and the ϕ azimuth angle is positive toward +X. The SWAP FOV is narrow in θ angular direction (~10°) and large in the ϕ angular direction (~276°).*

*Figure 4: Year long energy-time coincidence count rate spectrograms SWAP observations from 2012 January 1 through 2015 August 25 where the count rates have been corrected for changes with time in the detector efficiency using the Funsten et al.(2005).*

*Figure 5: Top: coincidence count rates for a given coarse-fine energy sweep with the coarse scan is in black, and the fine scan centered on the peak in the coarse scan is in blue. The right column is a zoom of the left column, and the middle and bottom panels show the location of the Sun in the polar (middle row) and azimuthal directions (bottom row).*

*Figure 6: The top panel shows the average coincidence rate squared divided by the average primary and secondary rate for the CEM detector gain flight calibrations. This ratio is a direct measure of the detection efficiency for the coincidence measurements. During these tests, the detector voltages were adjusted (color-coded in legend), and the observed count rates were recorded. The black bar across the top panel shows the duty cycle of the science observations, and the middle and bottom panels show the corresponding operational voltages for the science observations. The thick black line in the top panel beneath the colored lines is the final coincidence efficiency lookup table values corresponding to the operational values of the corresponding science observations.*

*Figure 7: Left: Color-coded coincidence count rate measurements) made in the laboratory with a beam energy ($E_{beam}$) of 1 keV with the beam at range of polar (θ) angles while stepping the observed energy ($E_{step}$) by adjusting the ESA voltage. Right: A SIMION® ion trajectory simulation of the energy-polar response for a 1 keV beam.*

*Figure 8 : Laboratory calibration of the SWAP instrument response (transmission) across the field-of-view in the polar and azimuthal direcitons with the electrostatic analyzer on and the Retarding Potential Analyzer (RPA) turned off. The laboratory beam was ~1 keV, and the*

*response is normalized to the value at 0° in θ and ϕ since other key laboratory calibration factors were determined at those angles.*

*Figure 9: The average transmission of the electrostatic analyzer response as a function of the azimuthal angle when the polar angle is between -1.5° and 1.5°.*

*Figure 10: Left: Location of the Sun in polar (y-axis) and azimuthal (x-axis) SWAP angular coordinates for a given coarse-fine energy sweep pair. Right: The location of the Sun in θ (black) and ϕ (red) as a function of time measured in seconds since the beginning of the energy sweep pair.*

*Figure 11: A histogram of the ratio of the semi-major and semi-minor axes corresponding to the amplitude in polar and azimuthal directions calculated for all the sweeps when the spacecraft was spinning from 2007 through 2015 August 25.*

*Figure 12: The origin of the circle that the Sun creates in θ (top) and ϕ (bottom) SWAP coordinates when the spacecraft is spinning. The black points correspond to individual sweeps where the fits had a reduced $\chi^2$ < 0.5, a Pearson correlation > 0.997 and Spearman correlation > 0.997. The red points are 3-day averages of the individual sweep results. The red points are used in the lookup table when fitting count rates while the spacecraft is spinning.*

*Figure 13: Overview flow diagram of the fitting procedure.*

*Figure 14: Flow diagram of the analytic model portion of the fit procedure. The reduced chi-square is given the symbol $\chi_r^2$. It is the $\chi^2$ divided by the number of degrees of freedom.*

*Figure 15: Flow diagram of the comprehensive model portion of the fit procedure.*

*Figure 16: The top row shows a close up of the count rates for a single coarse-fine pair of scans over the proton peaks. Each column has observations taken at different distances. The bottom two rows show the corresponding locations of the Sun in the polar (middle row) and azimuthal (bottom row) directions. Measurements from the fine scans are in black, and the coarse scan points are in purple. The analytic fit results are in red, and the comprehensive model fits are in blue. In the third column is an example recorded during hibernation when no attitude information is available.*

*Figure 17: In the left column each panel shows the $\chi_r^2$ as a function of percentage change away from the final fit (0%) values. These are results from a fit to the proton peak count rates measured in a single sweep. In the top panel the speed is varied, and similarly, the temperature and density are varied in the middle and bottom panels. The right column shows color distributions of the $\chi_r^2$ as a function of percentage change away from the final fit (0%) value for two different fit parameters. The speed and density are varied in the first panel. The speed and temperature and the density and temperature are varied in the second and third panels respectively.*

*Figure 18: Energy per charge-time corrected count rate spectrogram and corresponding solar wind speed, density and temperature time series for an active interval on the left and a quiet interval on the right.*

*Figure 19: Radial profiles of the solar wind proton density, temperature, speed, dynamic pressure and thermal pressure. The amplitude was fixed to the 1 AU value and the individual data points (black) beyond 22.2 AU were fit with a power law. The rotation averages are shown in orange.*

*Figure 20: Solar wind density (top), temperature (middle), and spacecraft latitude from 10 to 35 AU as a function of distance for both New Horizons (blue) and Voyager 2 (black). The SWAP measurements are at the cadence they were recorded. The individual SWAP count rates are always accumulated at 0.39 seconds, and a coarse-fine energy sweep pair is completed in 64 seconds. Often every sweep is not recorded. For example, during the long cruise intervals may have a one- or two-hour gap between sweeps; yet, the individual sweeps were still collected over a 64 second interval. During the Pluto sequence and annual checkout intervals the gap between the sweeps was often shorter. The Voyager 2 observations all have a 1 hour a cadence.*

*Figure 21: The running rotation averages of the solar wind speed from ACE (black), STEREO A (red), and STEREO B (blue) propagated to the radial distance of New Horizons (orange). The propagation time is determined using the running rotation average speed. The New Horizons speeds are at the cadence they were collected and not averaged. The 1 AU segments occur when the longitude separation with New Horizons (bottom panel) was with in 70° longitude.*

*Figure 22: The top two panels are time series of the solar wind speed and temperature at the sweep cadence recorded. The bottom two panels are running averages of the speed (2-day window) and temperature (7-day window) with color-coding. The color-coding for the speed based on the 2-day running speed-time slope. The color is orange when the slope is $> 3.17 \times 10^{-5}$ km s$^{-2}$ and blue when the slope is $<-3.17 \times 10^{-5}$ km s$^{-2}$. The color for the temperature is based on a 7-day average temperature-speed slope. When the temperature-speed slope is $>1.7x^{-4}$ K s km$^{-1}$, the color is pink and the color is green when the temperature-speed slope is $<1.7 \times 10^{-4}$ K s km$^{-1}$.*

*Figure 23: Color distribution plots of the normalized temperature versus the solar wind speed. The top row shows observations fro Helios, ACE, and Ulysses at different distance ranges (adapted from Figures 11-13 in Elliott et al. (2012). In the bottom row are observations from NH-SWAP. The left panel shows all the observations starting in 2012 through late August 25, 2015. The middle and right panels show the same measurements divided in half by distance range. Linear fit is shown in pink. The Spearman (SCorr.) and Pearson (PCorr.) correlation coeffcents are labeled below the distance range.*

*Figure 24: The panels from top to bottom are periodograms of 1 AU ACE speed, temperature, and density measurements. The Lomb-Scargle (L-S) periodograms are shown in orange with the y-axis on the right. The y-axis for the Discrete Fourier Transform (DFT) (light blue) and the cleaned DFT (black) periodograms is on the left.*

*Figure 25: This figure has the same format as Figure 24, but here the periodograms are derived from SWAP data from 21.96 to 25.80 AU.*

*Figure 26: Periodograms in the same format as Figure 25, but for the distance range from 25.80 to 29.54 AU.*

*Figure 27: Periodograms in the same format as Figure 25, but for the distance range from 29.54 to 33.26 AU.*

*Figure A1: Simulated count rates using the analytic model, which simplifies the integration are shown as thin lines in lighter colors. The thicker and darker curves are full numerical integration using only the SWAP effective area and the field-of-view. On the left the temperature is varied and on the right the density is varied.*

*Figure A2: Analytic and comprehensive temperatures and densities compared to one another and piece-wise fit. The data spans from 2008 October 10 to 2014 July 7.*

TABLES:

*Table 1: Table of exponents determined from power law fits to solar wind parameters as a function of distance.*

| Quantity | Voyager 2 1-30 AU \|Lat.\| <4° (Wang & Richardson 2004) | Voyager 2 1-45 AU \|Lat.\| <4° (Richardson et al. 1996) | Ulysses Polar Holes 1-5.4 AU \|Lat.\| >36° (McComas et al. 2000) | Helios and Ulysses 0.29-5.4 AU Lat.\| <36° (Elliott et al. 2012) | SWAP 22.2-33.3 AU \|Lat.\| <6° |
|---|---|---|---|---|---|
| V | 0.014 | N/A | N/A | N/A | -0.010 |
| n | -1.97 | -1.93 | -2.00 | -2.25 | -1.77 |
| T | -0.62 | -0.46 | -1.02 | -0.72 | -0.745 |

*Table 2: A table of slopes and y-intercepts based on linear fits for given temperature radial normalization exponents and distance ranges.*

| Exponent | Distance Range [AU] | m [Ks/km] | b [K] |
|---|---|---|---|
| 1.0 | 21.96 - 33.26 | 3091 | $-8.93\times10^5$ |
|  | 21.96 - 27.68 | 1817 | $-5.19\times10^5$ |
|  | 27.68 - 33.26 | 3320 | $-9.69\times10^5$ |
| 0.745 | 21.96 - 33.26 | 1281 | $-3.71\times10^5$ |
|  | 21.96 - 27.68 | 809 | $-2.32\times10^5$ |
|  | 27.68 - 33.26 | 1366 | $-3.99\times10^5$ |

FIGURES:

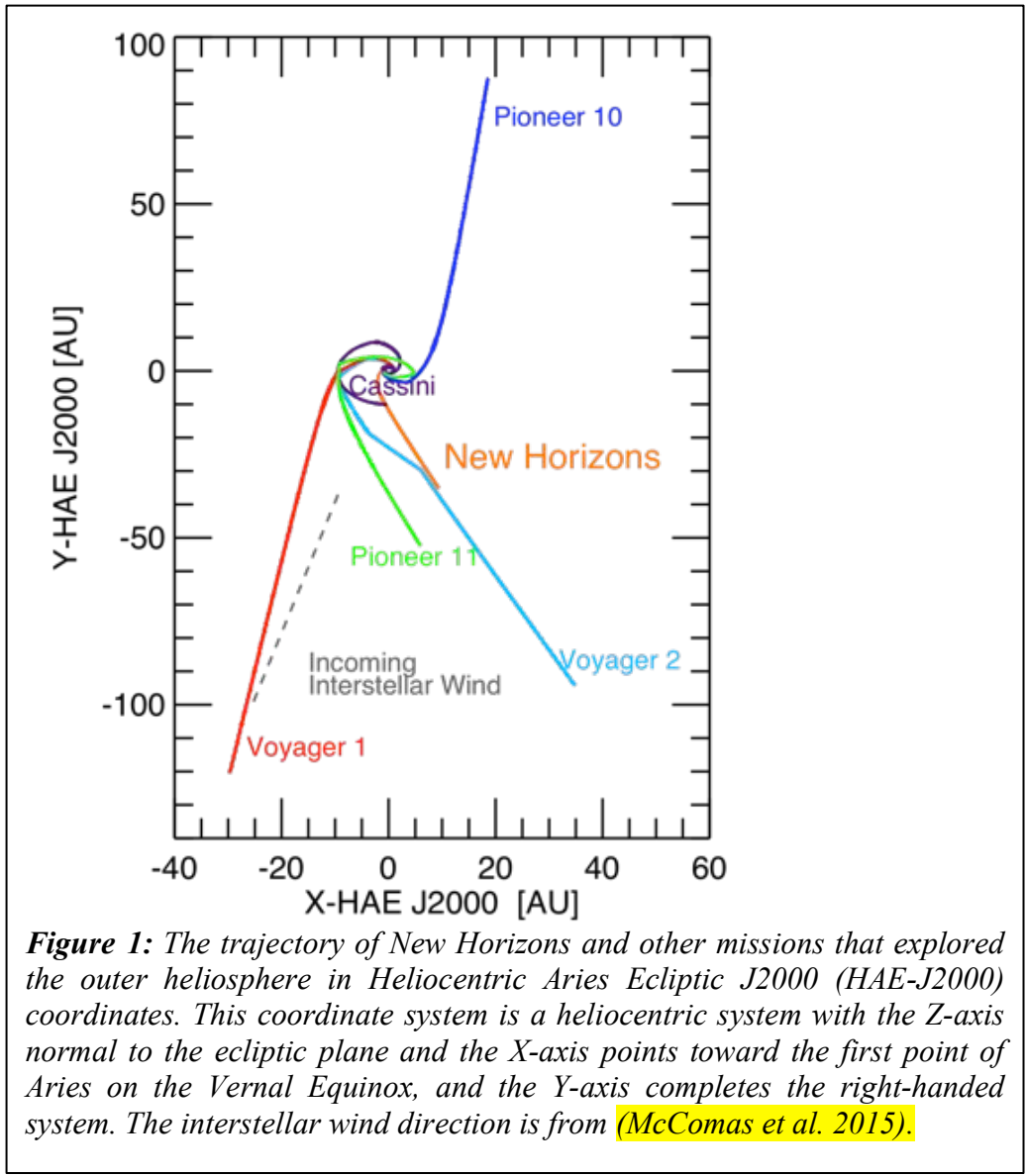

*Figure 1:* The trajectory of New Horizons and other missions that explored the outer heliosphere in Heliocentric Aries Ecliptic J2000 (HAE-J2000) coordinates. This coordinate system is a heliocentric system with the Z-axis normal to the ecliptic plane and the X-axis points toward the first point of Aries on the Vernal Equinox, and the Y-axis completes the right-handed system. The interstellar wind direction is from (McComas et al. 2015).

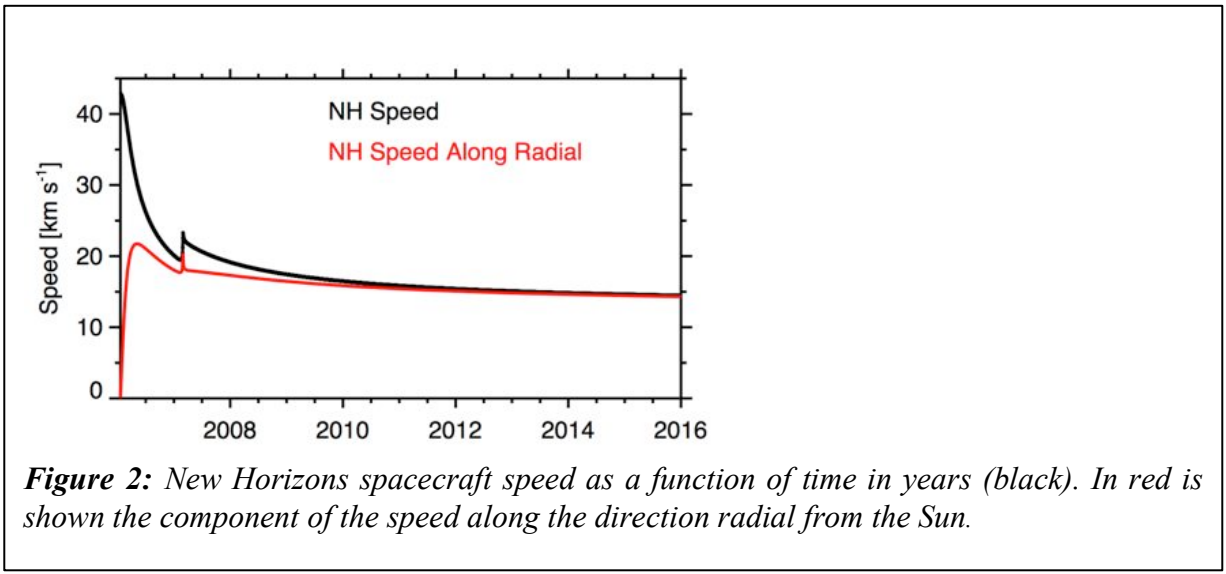

*Figure 2:* New Horizons spacecraft speed as a function of time in years (black). In red is shown the component of the speed along the direction radial from the Sun.

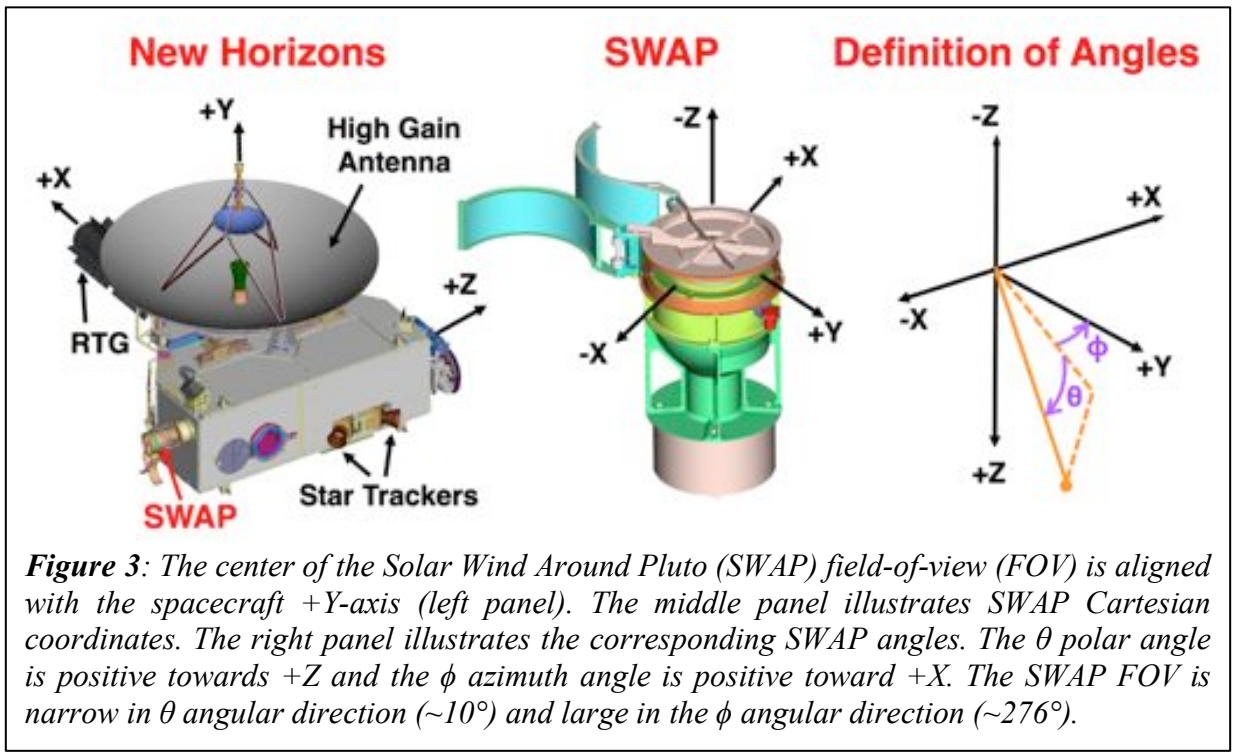

*Figure 3*: The center of the Solar Wind Around Pluto (SWAP) field-of-view (FOV) is aligned with the spacecraft +Y-axis (left panel). The middle panel illustrates SWAP Cartesian coordinates. The right panel illustrates the corresponding SWAP angles. The θ polar angle is positive towards +Z and the ϕ azimuth angle is positive toward +X. The SWAP FOV is narrow in θ angular direction (~10°) and large in the ϕ angular direction (~276°).

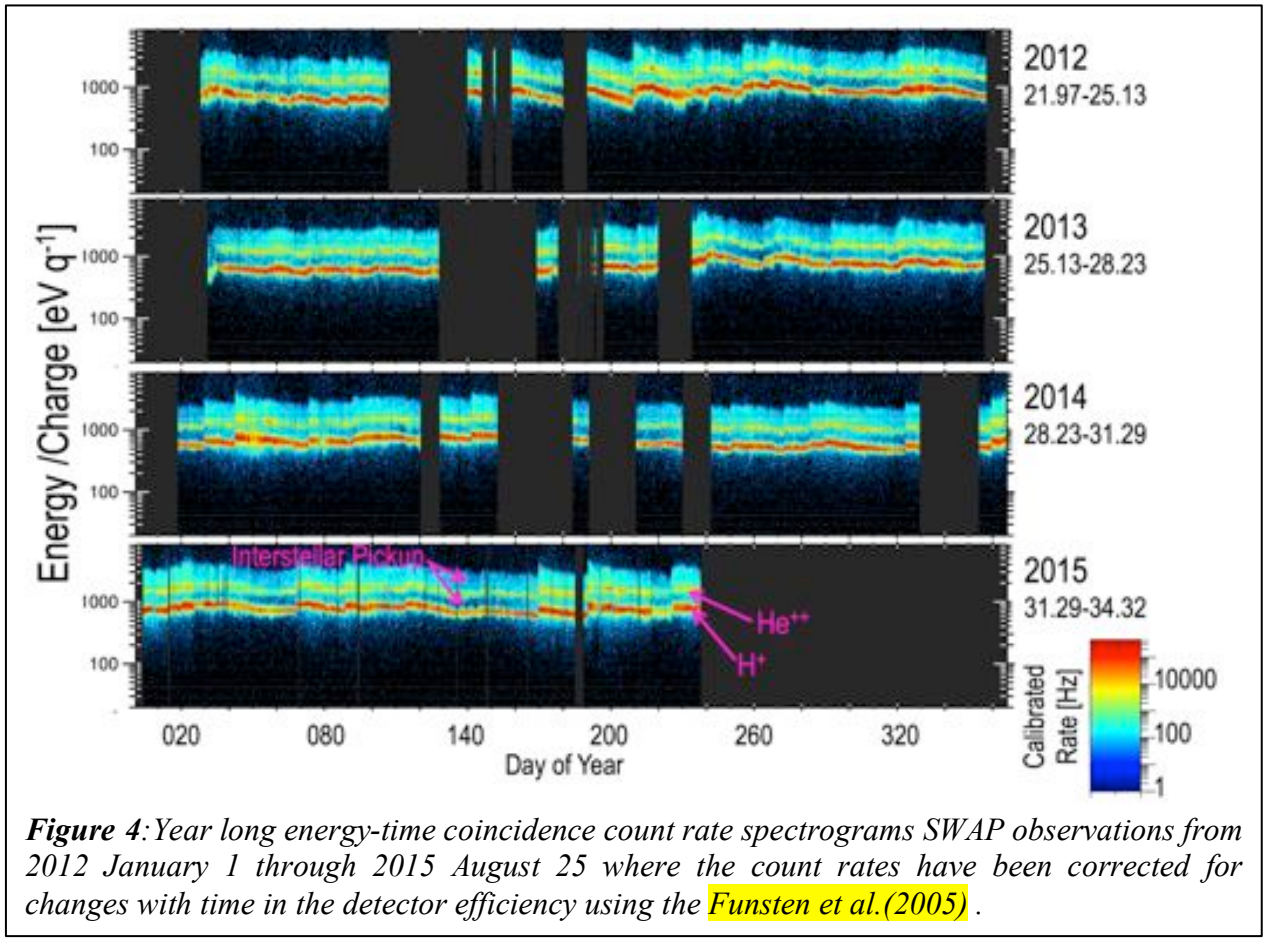

***Figure 4***: Year long energy-time coincidence count rate spectrograms SWAP observations from 2012 January 1 through 2015 August 25 where the count rates have been corrected for changes with time in the detector efficiency using the Funsten et al.(2005) .

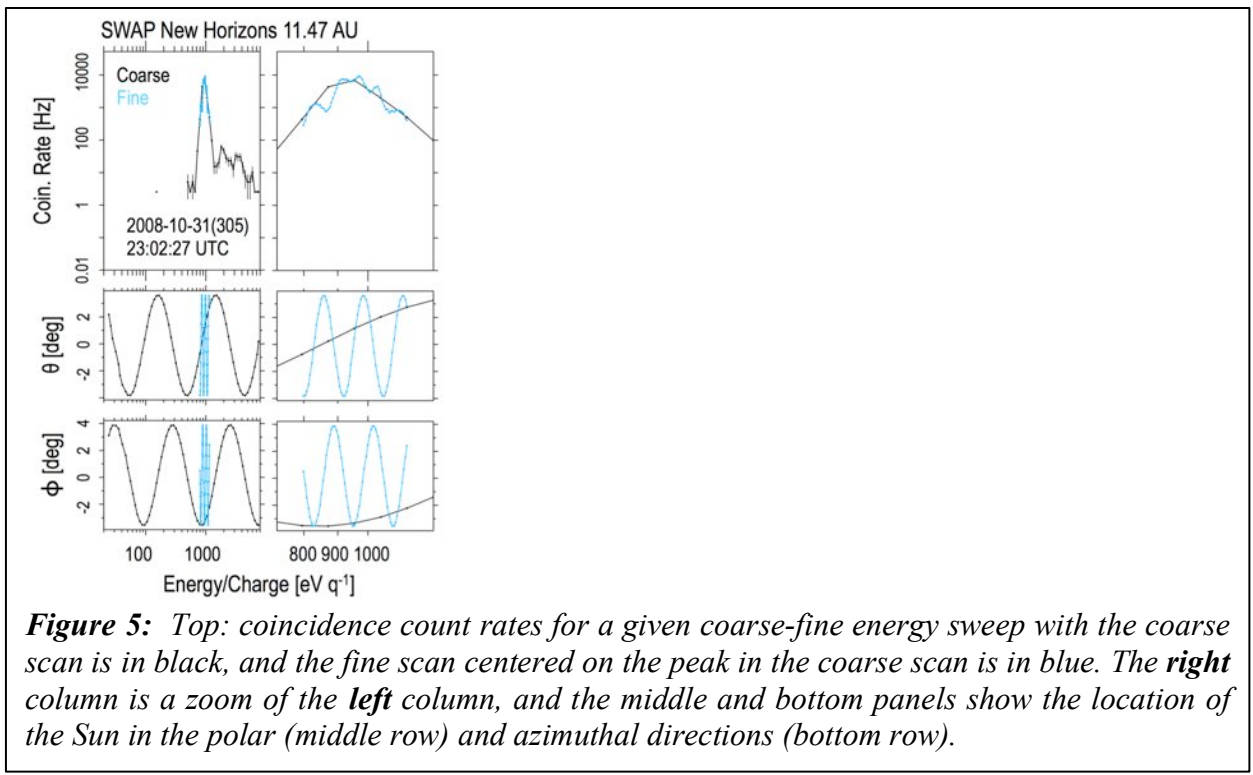

***Figure 5:*** Top: coincidence count rates for a given coarse-fine energy sweep with the coarse scan is in black, and the fine scan centered on the peak in the coarse scan is in blue. The **right** column is a zoom of the **left** column, and the middle and bottom panels show the location of the Sun in the polar (middle row) and azimuthal directions (bottom row).

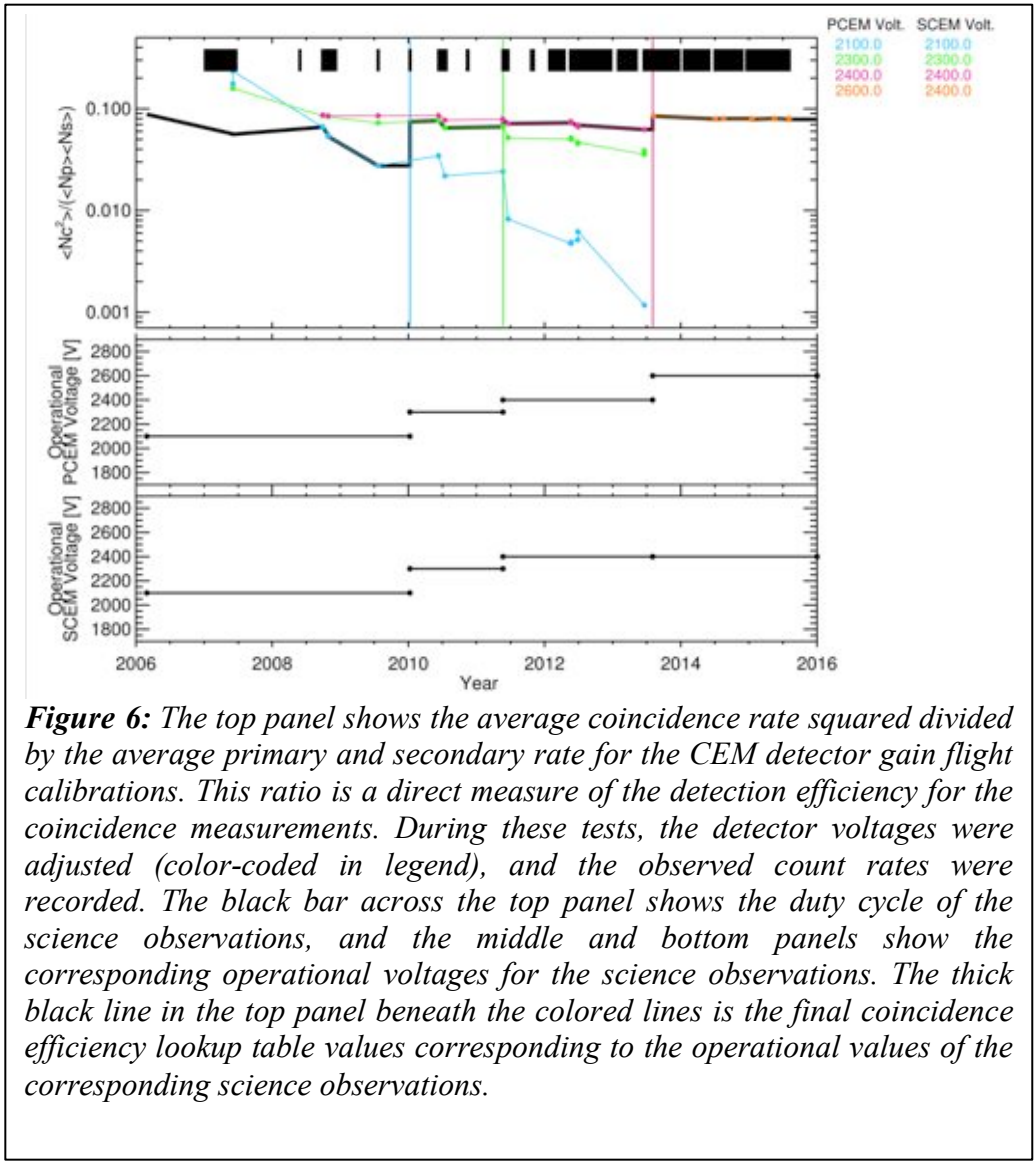

*Figure 6:* *The top panel shows the average coincidence rate squared divided by the average primary and secondary rate for the CEM detector gain flight calibrations. This ratio is a direct measure of the detection efficiency for the coincidence measurements. During these tests, the detector voltages were adjusted (color-coded in legend), and the observed count rates were recorded. The black bar across the top panel shows the duty cycle of the science observations, and the middle and bottom panels show the corresponding operational voltages for the science observations. The thick black line in the top panel beneath the colored lines is the final coincidence efficiency lookup table values corresponding to the operational values of the corresponding science observations.*

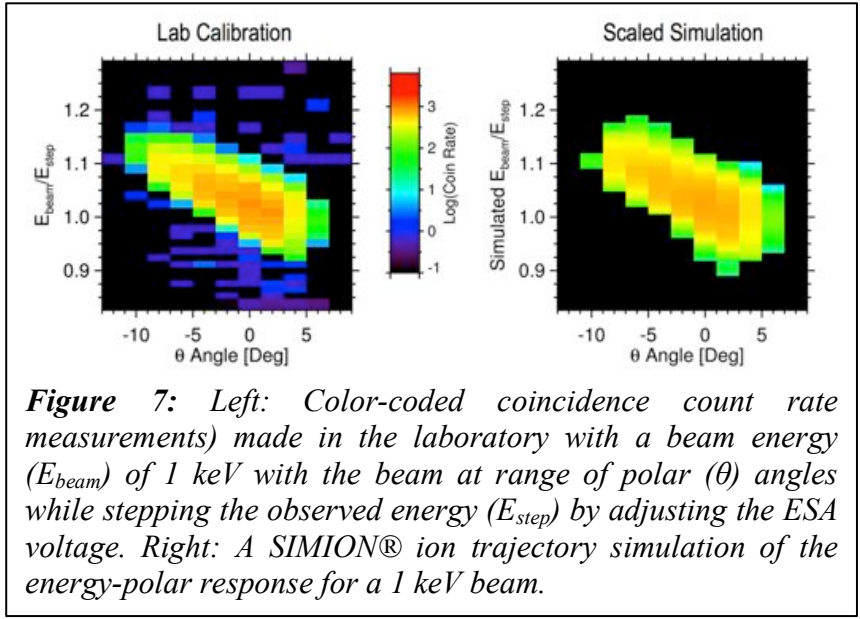

*Figure 7:* Left: Color-coded coincidence count rate measurements) made in the laboratory with a beam energy ($E_{beam}$) of 1 keV with the beam at range of polar ($\theta$) angles while stepping the observed energy ($E_{step}$) by adjusting the ESA voltage. Right: A SIMION® ion trajectory simulation of the energy-polar response for a 1 keV beam.

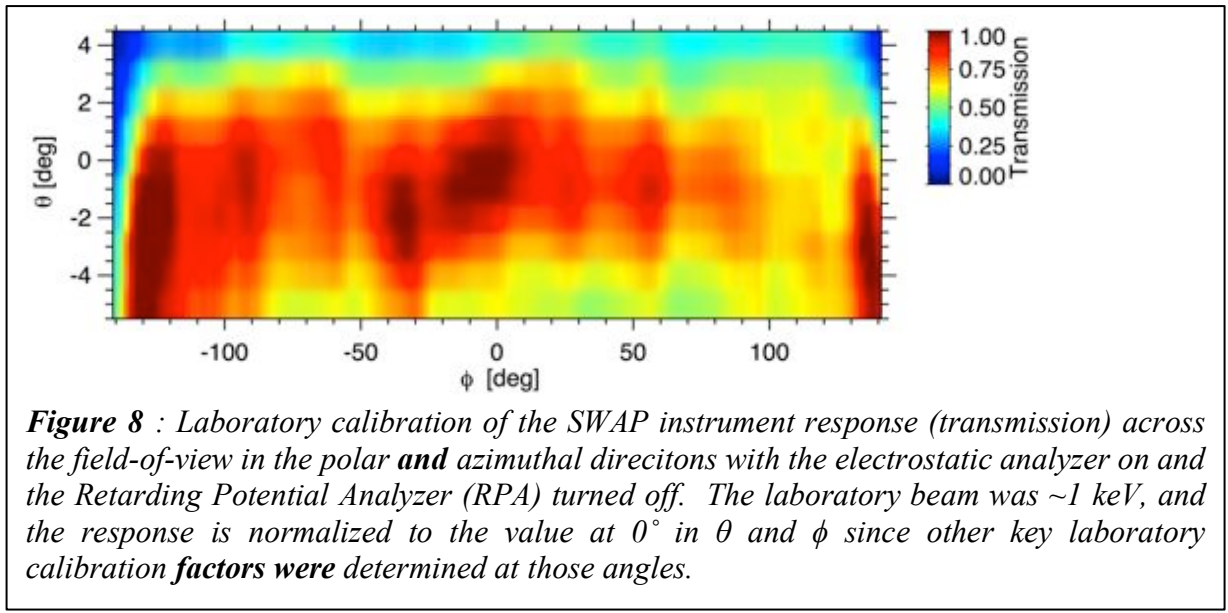

*Figure 8* : Laboratory calibration of the SWAP instrument response (transmission) across the field-of-view in the polar **and** azimuthal direcitons with the electrostatic analyzer on and the Retarding Potential Analyzer (RPA) turned off. The laboratory beam was ~1 keV, and the response is normalized to the value at 0˚ in $\theta$ and $\phi$ since other key laboratory calibration **factors were** determined at those angles.

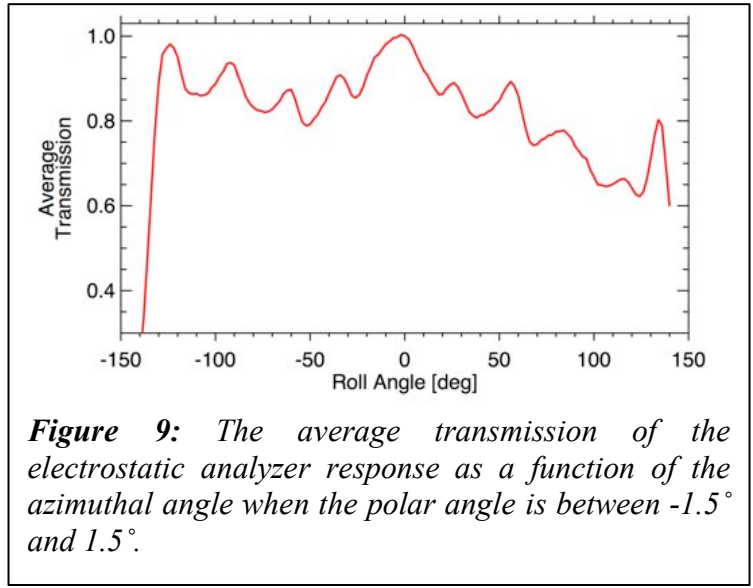

*Figure 9: The average transmission of the electrostatic analyzer response as a function of the azimuthal angle when the polar angle is between -1.5˚ and 1.5˚.*

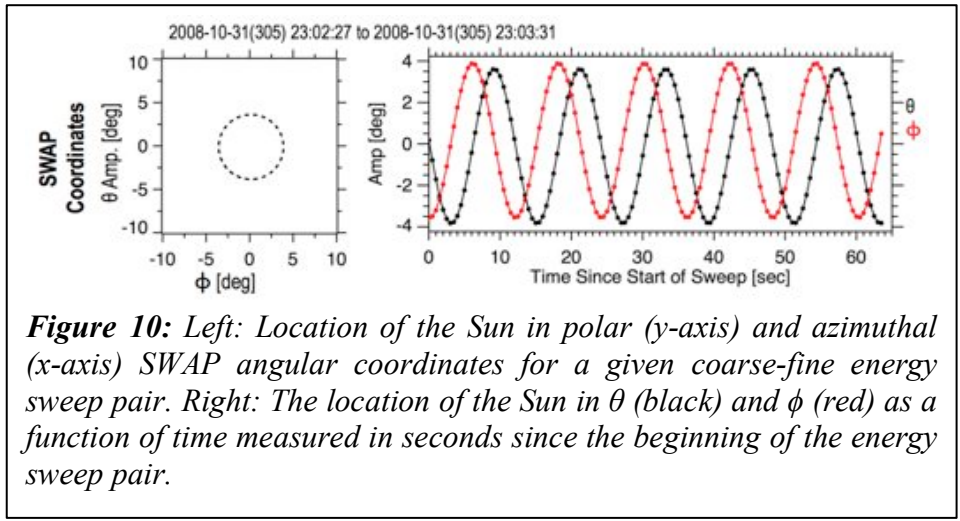

*Figure 10: Left: Location of the Sun in polar (y-axis) and azimuthal (x-axis) SWAP angular coordinates for a given coarse-fine energy sweep pair. Right: The location of the Sun in θ (black) and ϕ (red) as a function of time measured in seconds since the beginning of the energy sweep pair.*

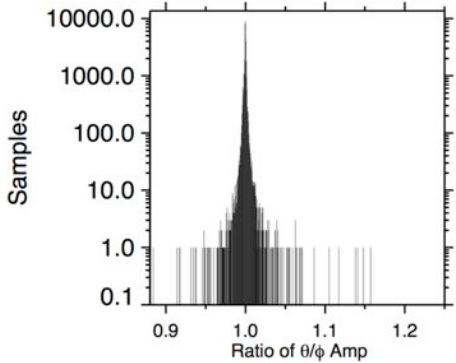

*Figure 11:* *A histogram of the ratio of the semi-major and semi-minor axes corresponding to the amplitude in polar and azimuthal directions calculated for all the sweeps when the spacecraft was spinning from 2007 through 2015 August 25.*

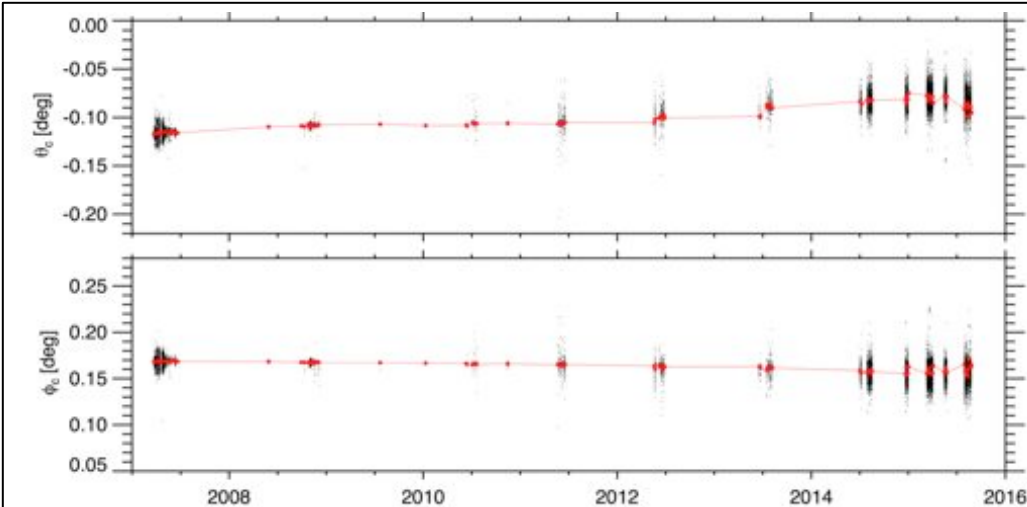

*Figure 12:* *The origin of the circle that the Sun creates in θ (top) and ϕ (bottom) SWAP coordinates when the spacecraft is spinning. The black points correspond to individual sweeps where the fits had a reduced $\chi^2$ < 0.5, a Pearson correlation > 0.997 and Spearman correlation > 0.997. The red points are 3-day averages of the individual sweep results. The red points are used in the lookup table when fitting count rates while the spacecraft is spinning.*

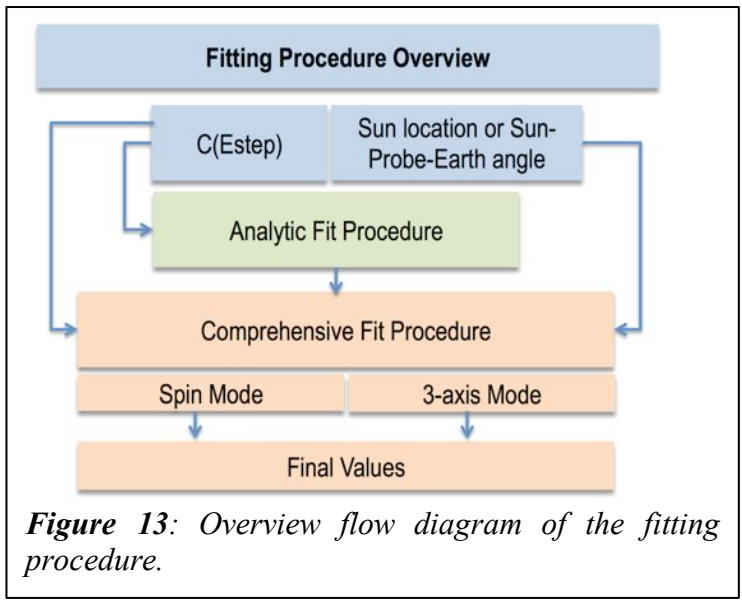

*Figure 13: Overview flow diagram of the fitting procedure.*

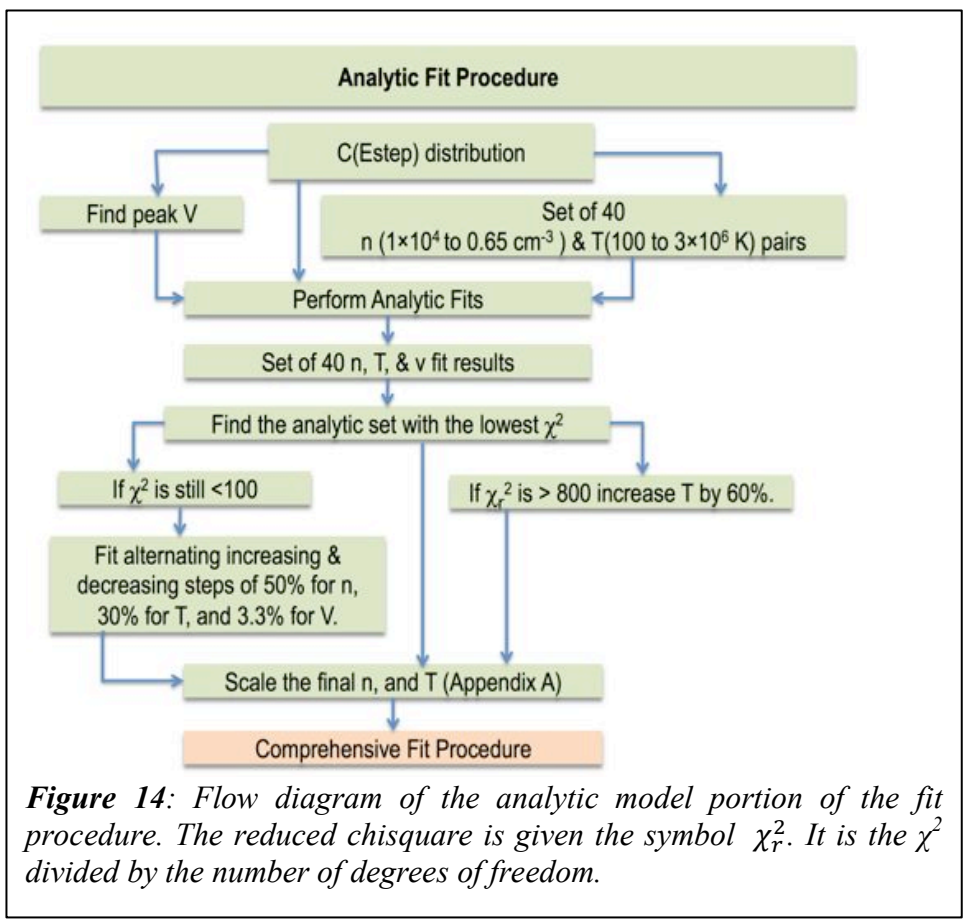

*Figure 14: Flow diagram of the analytic model portion of the fit procedure. The reduced chisquare is given the symbol $\chi_r^2$. It is the $\chi^2$ divided by the number of degrees of freedom.*

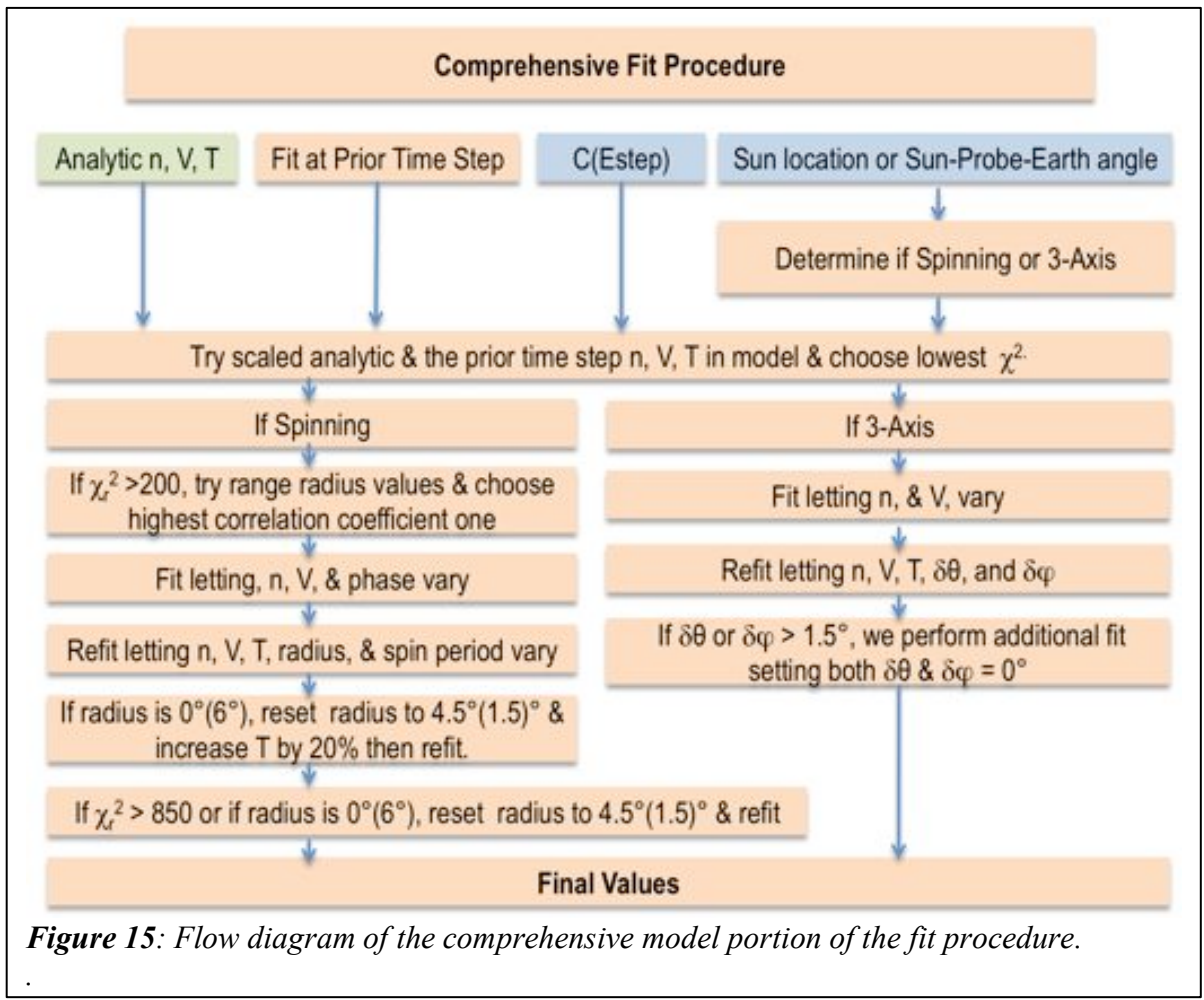

***Figure 15***: *Flow diagram of the comprehensive model portion of the fit procedure.*
.

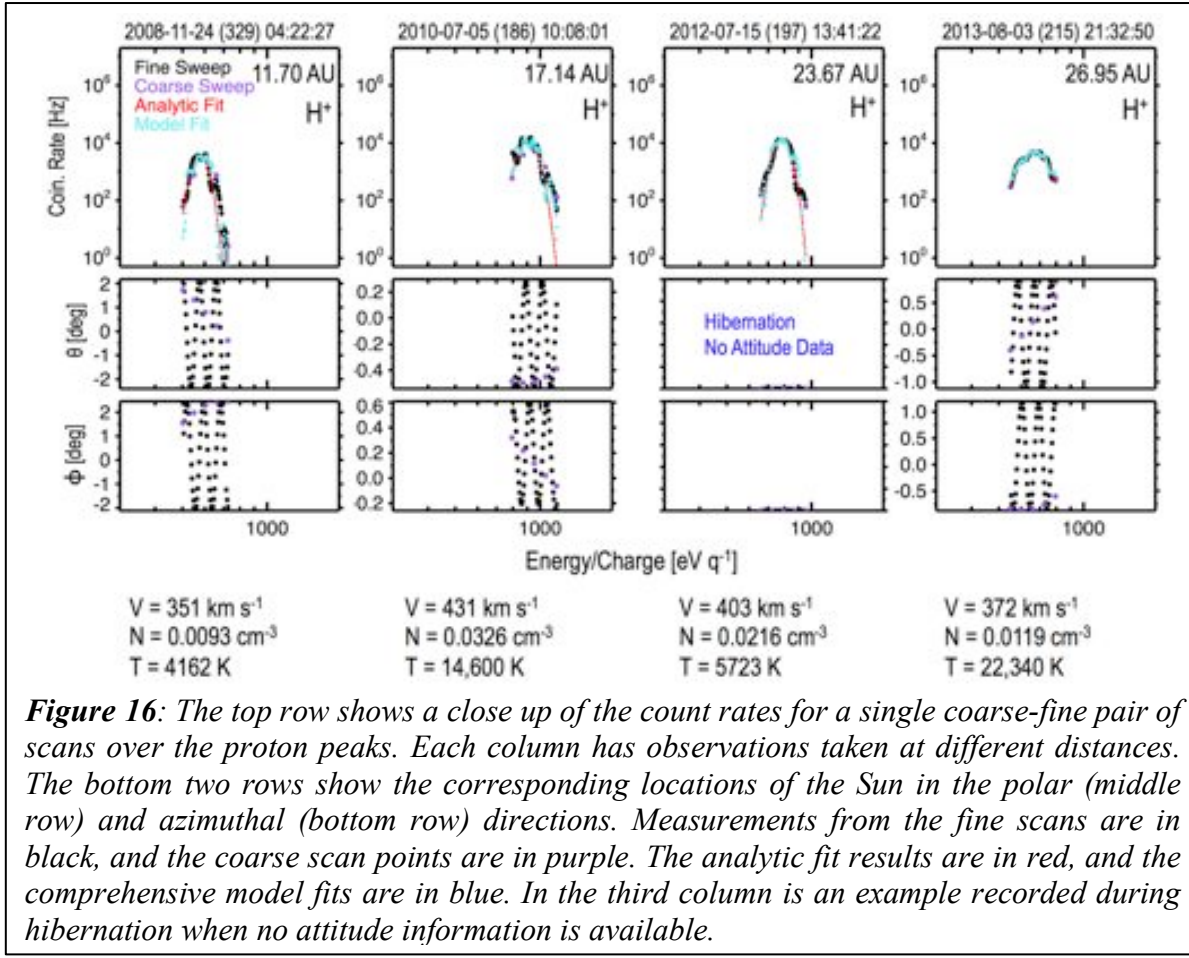

*Figure 16*: *The top row shows a close up of the count rates for a single coarse-fine pair of scans over the proton peaks. Each column has observations taken at different distances. The bottom two rows show the corresponding locations of the Sun in the polar (middle row) and azimuthal (bottom row) directions. Measurements from the fine scans are in black, and the coarse scan points are in purple. The analytic fit results are in red, and the comprehensive model fits are in blue. In the third column is an example recorded during hibernation when no attitude information is available.*

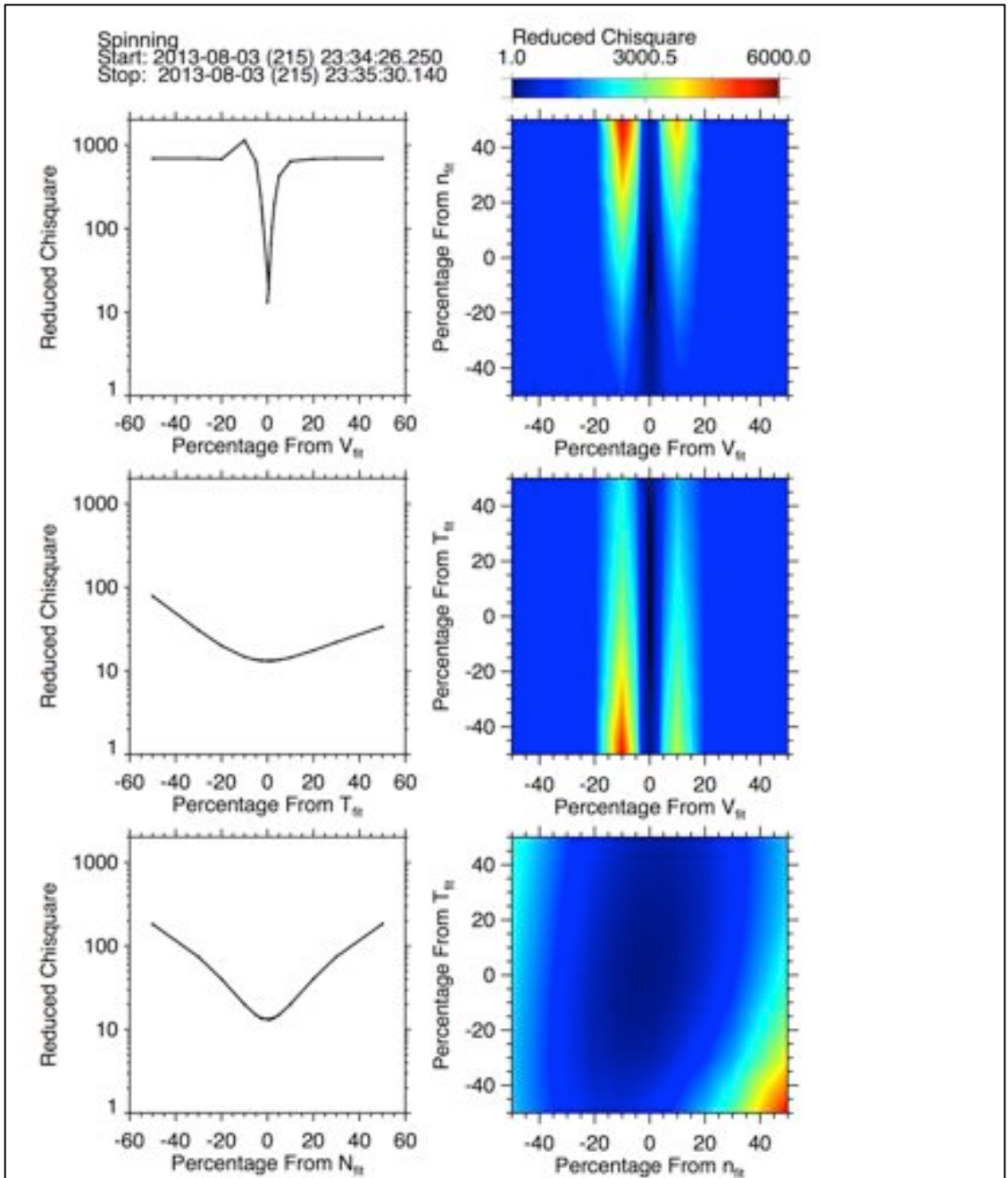

***Figure 17***: *In the left column each panel shows the $\chi_r^2$ as a function of percentage change away from the final fit (0%) values. These are results from a fit to the proton peak count rates measured in a single sweep. In the top panel the speed is varied, and similarly, the temperature and density are varied in the middle and bottom panels. The right column shows color distributions of the $\chi_r^2$ as a function of percentage change away from the final fit (0%) value for two different fit parameters. The speed and density are varied in the first panel. The speed and temperature and the density and temperature are varied in the second and third panels respectively.*

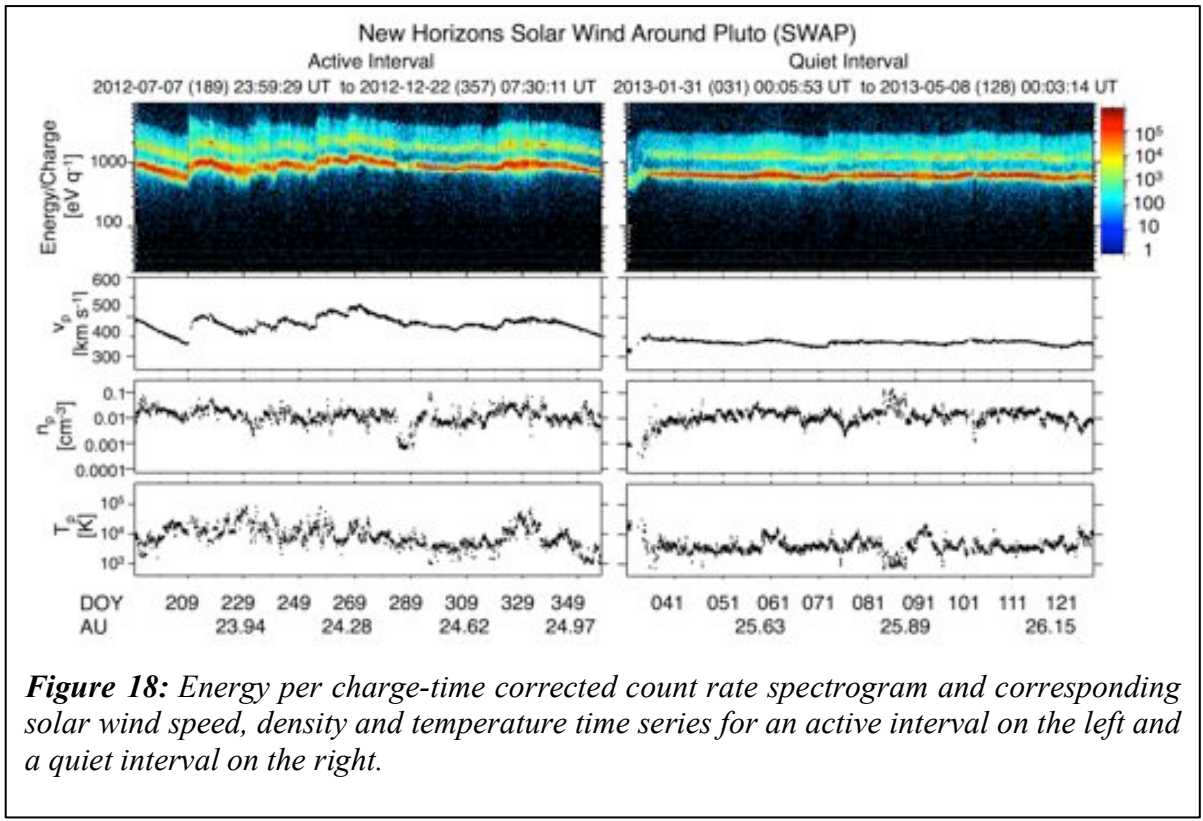

*Figure 18:* *Energy per charge-time corrected count rate spectrogram and corresponding solar wind speed, density and temperature time series for an active interval on the left and a quiet interval on the right.*

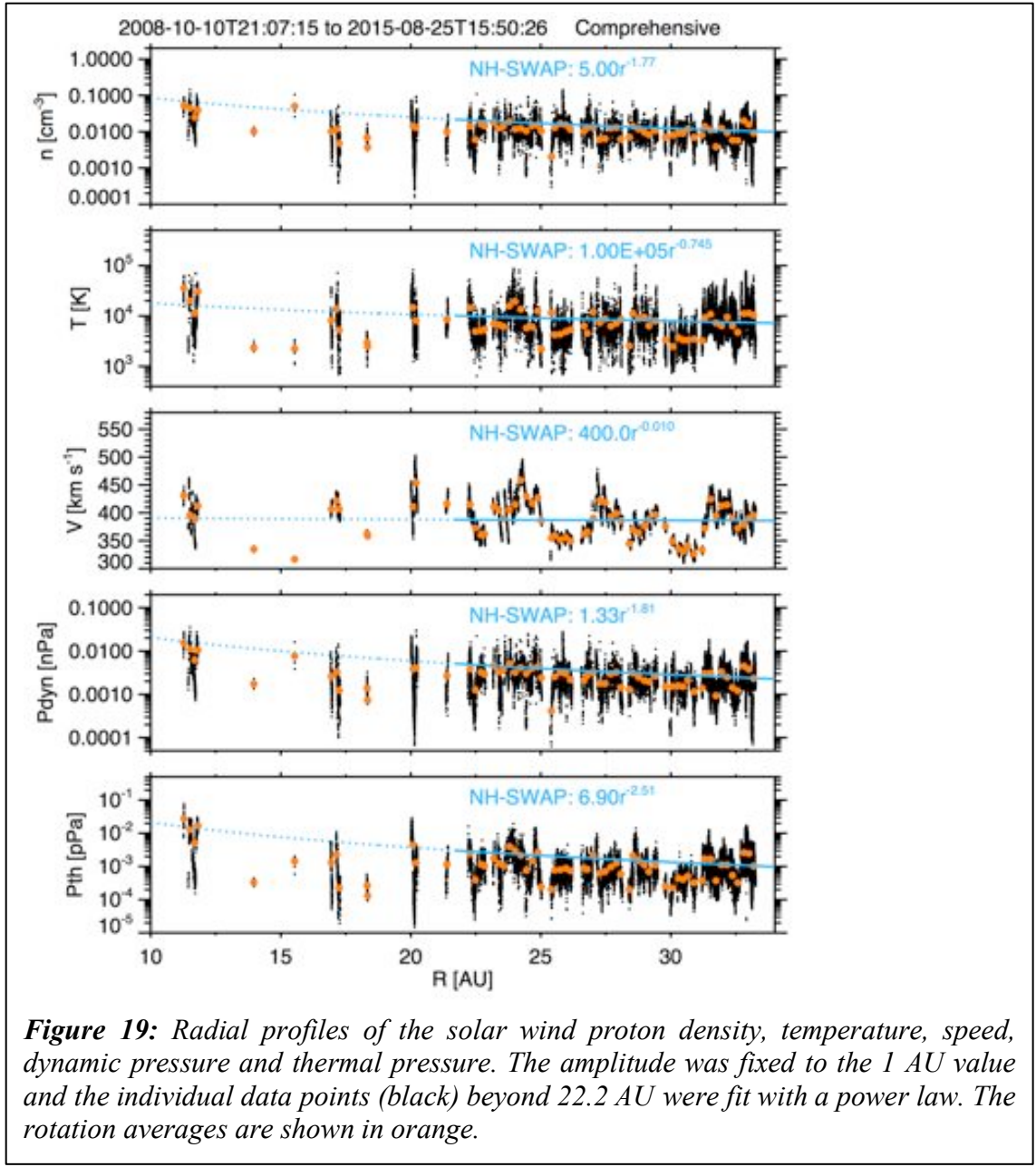

*Figure 19: Radial profiles of the solar wind proton density, temperature, speed, dynamic pressure and thermal pressure. The amplitude was fixed to the 1 AU value and the individual data points (black) beyond 22.2 AU were fit with a power law. The rotation averages are shown in orange.*

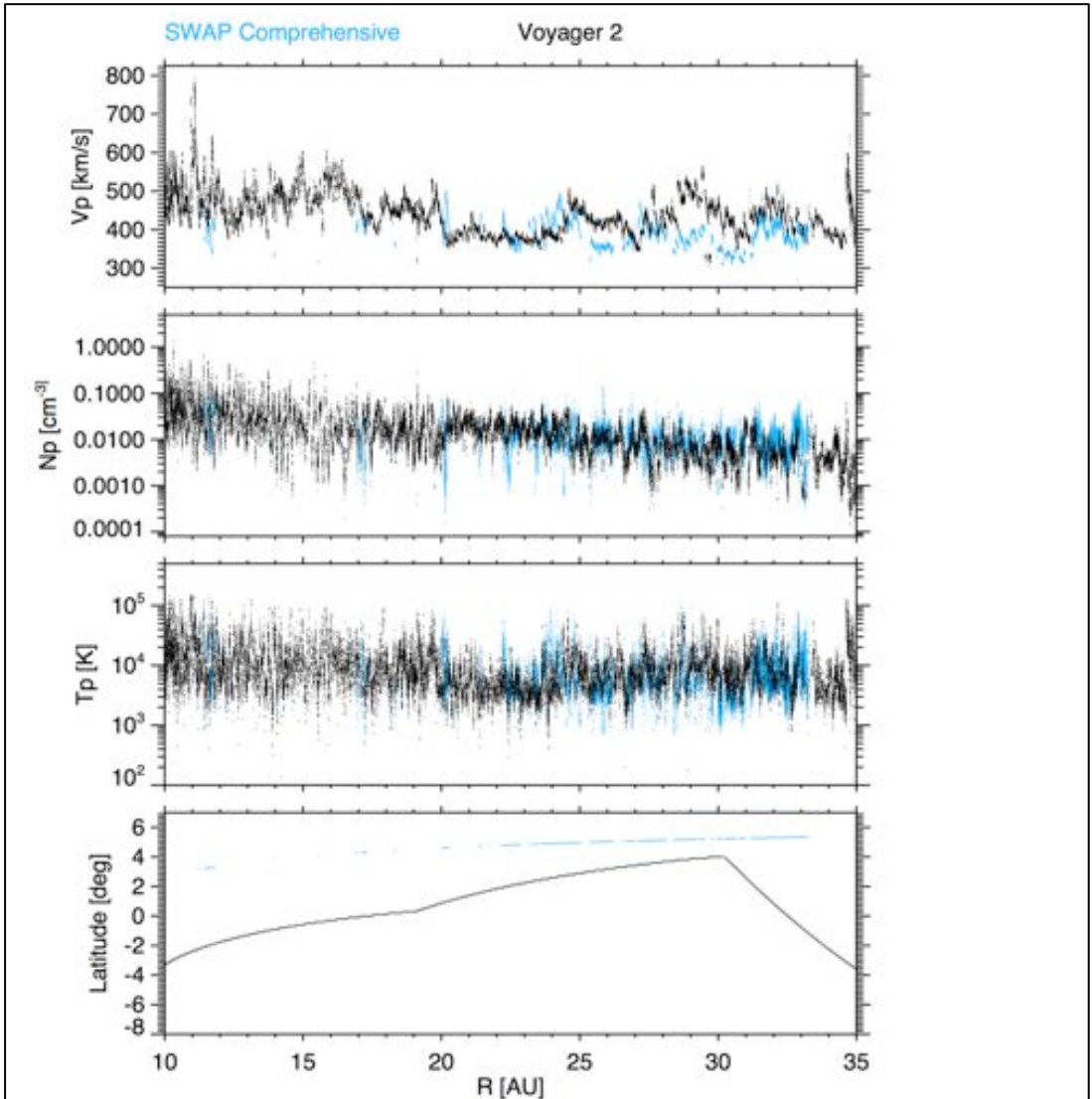

*Figure 20:* *Solar wind density (top), temperature (middle), and spacecraft latitude from 10 to 35 AU as a function of distance for both New Horizons (blue) and Voyager 2 (black). The SWAP measurements are at the cadence they were recorded. The individual SWAP count rates are always accumulated at 0.39 seconds, and a coarse-fine energy sweep pair is completed in 64 seconds. Often every sweep is not **recorded**. For example, during the long cruise intervals may have a **one- or two-hour** gap between sweeps; yet, the individual sweeps were still collected over a 64 second interval. During the Pluto sequence and annual checkout intervals the **gap** between the sweeps was often shorter. The Voyager 2 observations all have **a** 1 hour a cadence.*

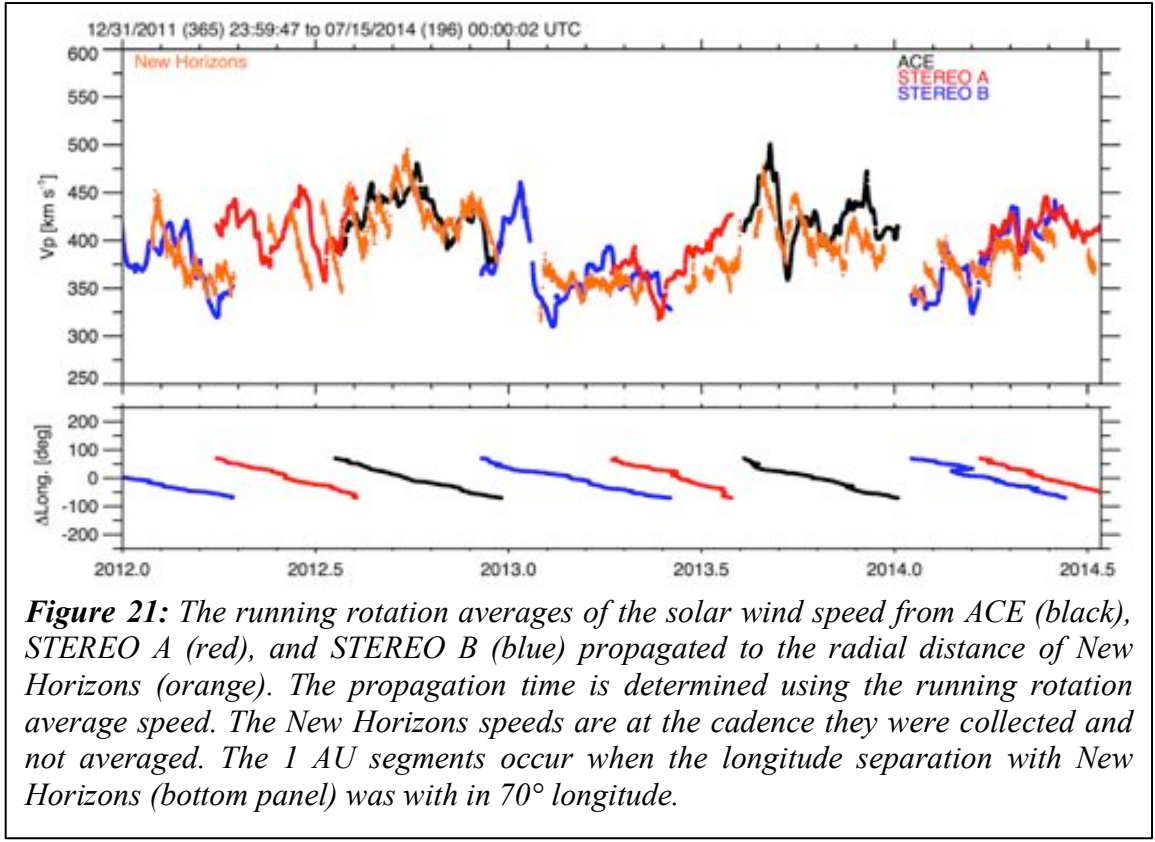

*Figure 21:* *The running rotation averages of the solar wind speed from ACE (black), STEREO A (red), and STEREO B (blue) propagated to the radial distance of New Horizons (orange). The propagation time is determined using the running rotation average speed. The New Horizons speeds are at the cadence they were collected and not averaged. The 1 AU segments occur when the longitude separation with New Horizons (bottom panel) was with in 70° longitude.*

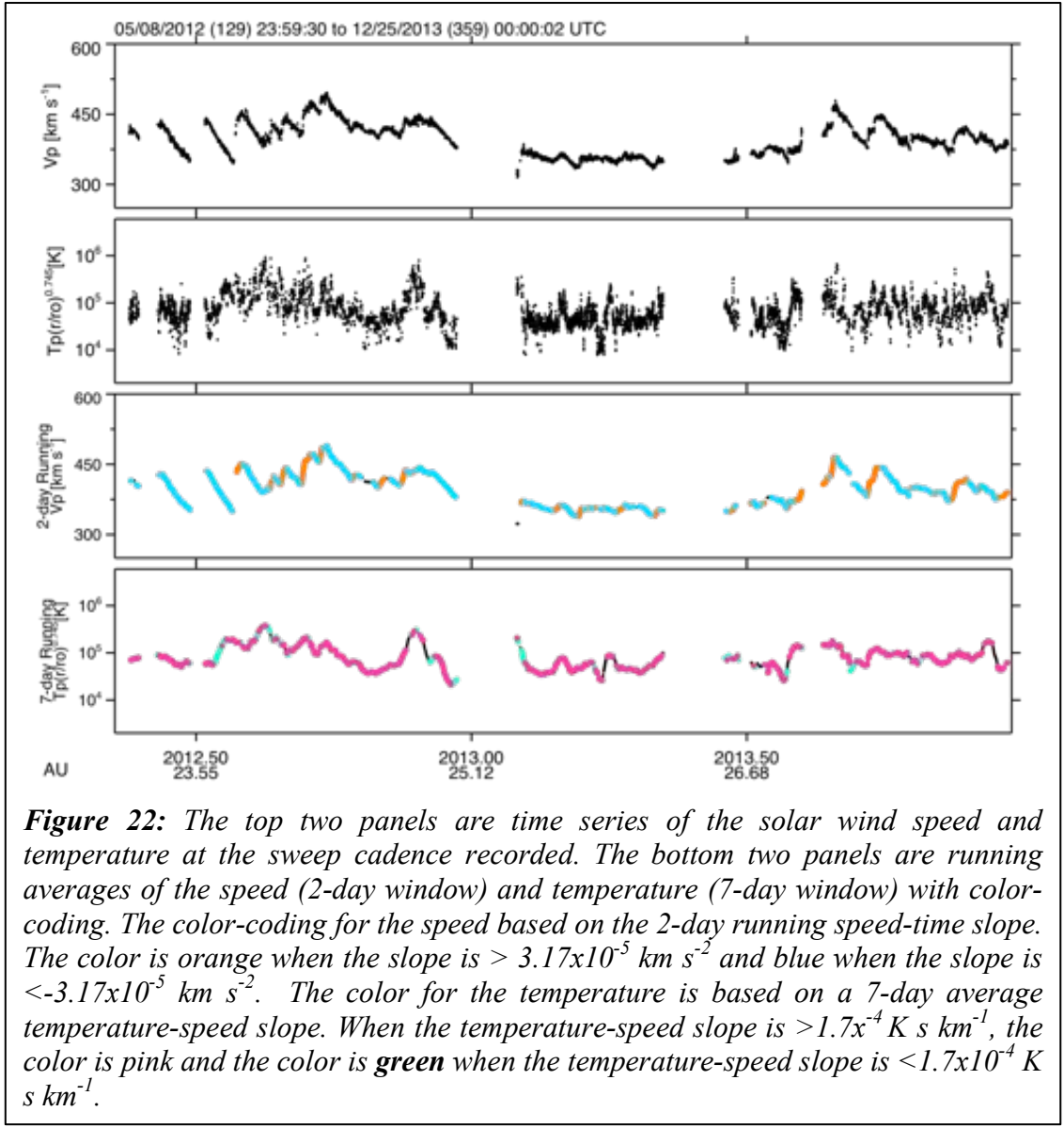

*Figure 22:* *The top two panels are time series of the solar wind speed and temperature at the sweep cadence recorded. The bottom two panels are running averages of the speed (2-day window) and temperature (7-day window) with color-coding. The color-coding for the speed based on the 2-day running speed-time slope. The color is orange when the slope is $> 3.17 \times 10^{-5}$ km s$^{-2}$ and blue when the slope is $<-3.17 \times 10^{-5}$ km s$^{-2}$. The color for the temperature is based on a 7-day average temperature-speed slope. When the temperature-speed slope is $>1.7 \times 10^{-4}$ K s km$^{-1}$, the color is pink and the color is **green** when the temperature-speed slope is $<1.7 \times 10^{-4}$ K s km$^{-1}$.*

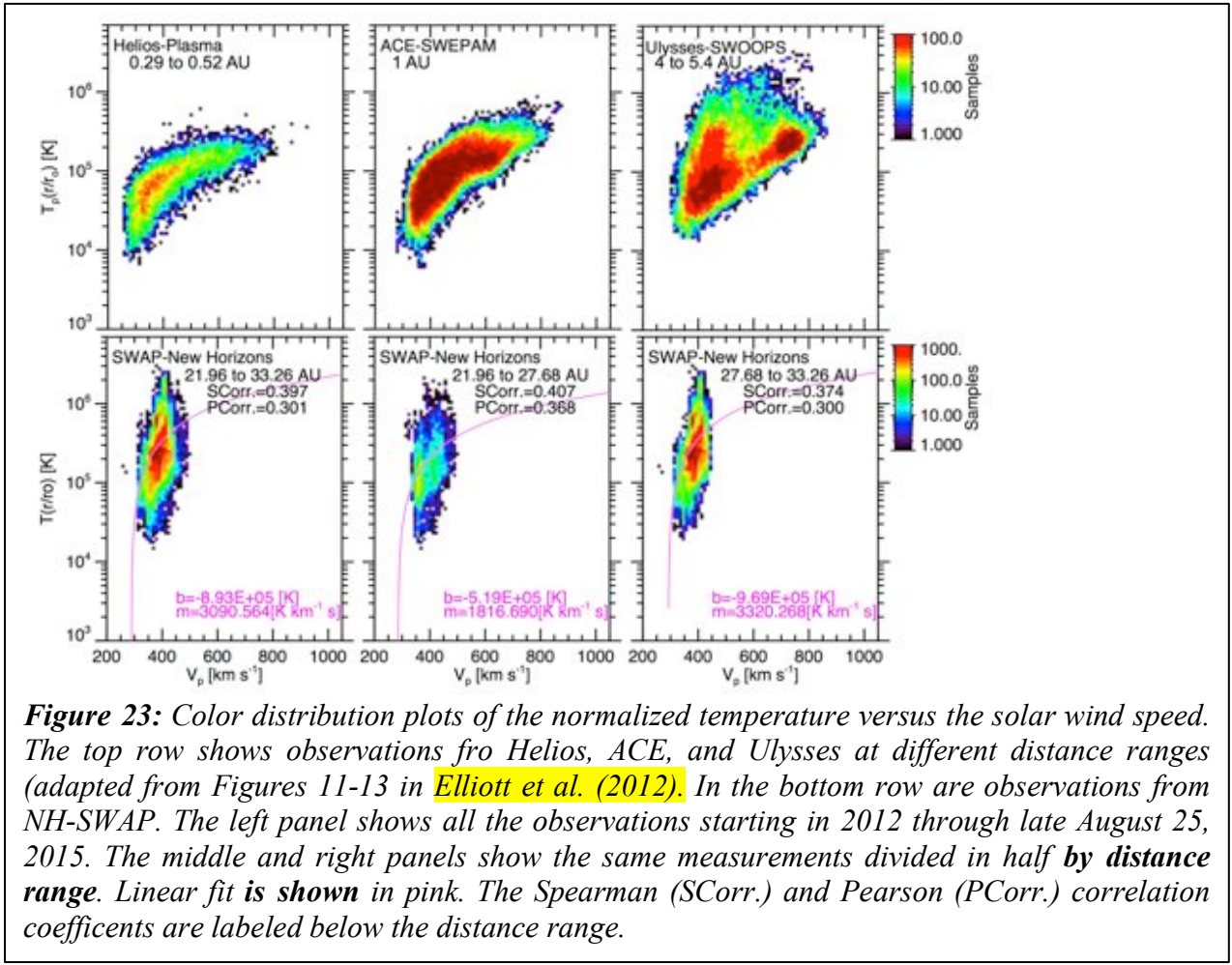

*Figure 23:* Color distribution plots of the normalized temperature versus the solar wind speed. The top row shows observations fro Helios, ACE, and Ulysses at different distance ranges (adapted from Figures 11-13 in Elliott et al. (2012). In the bottom row are observations from NH-SWAP. The left panel shows all the observations starting in 2012 through late August 25, 2015. The middle and right panels show the same measurements divided in half **by distance range**. Linear fit **is shown** in pink. The Spearman (SCorr.) and Pearson (PCorr.) correlation coefficents are labeled below the distance range.

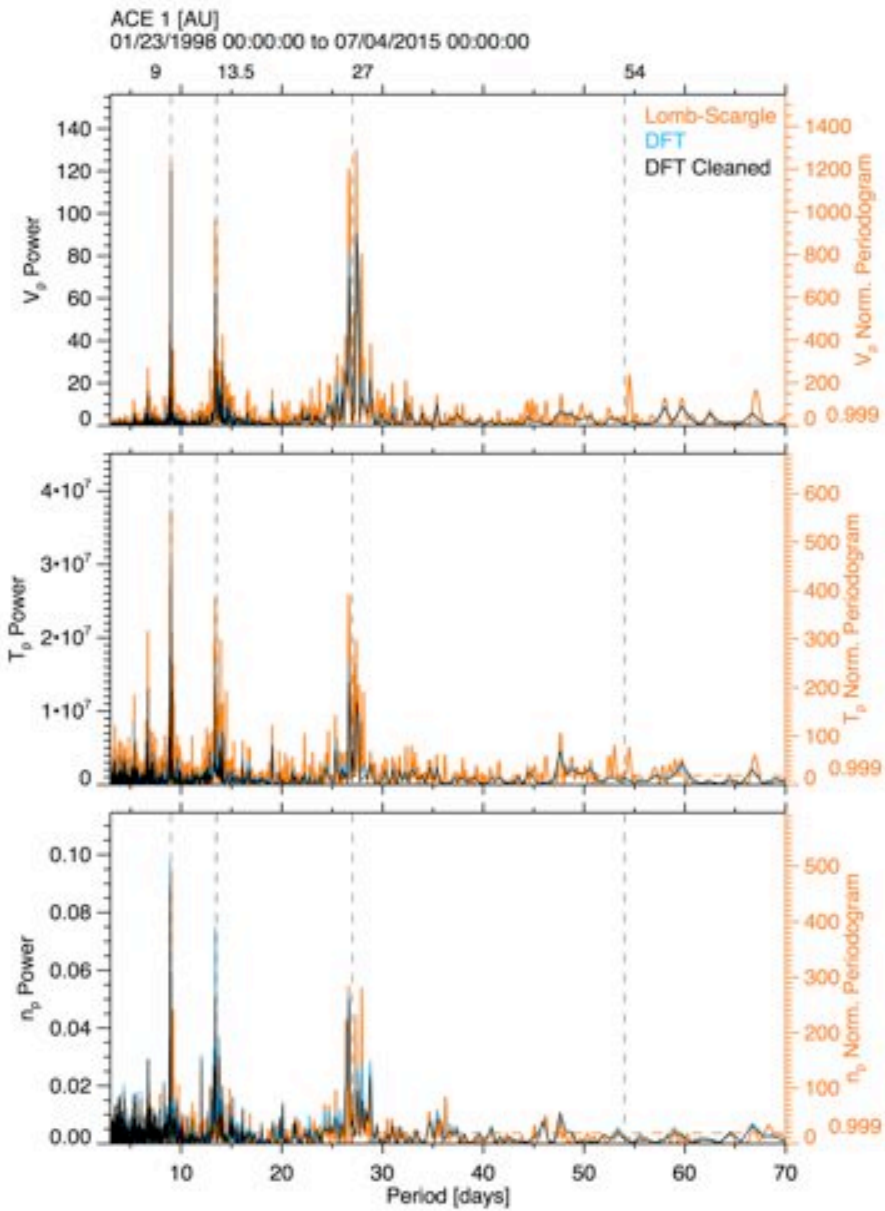

*Figure 24: The panels from top to bottom are periodograms of 1 AU ACE speed, temperature, and density measurements. The Lomb-Scargle (L-S) periodograms are shown in orange with the y-axis on the right. The y-axis for the Discrete Fourier Transform (DFT) (light blue) and the cleaned DFT (black) periodograms is on the **left**.*

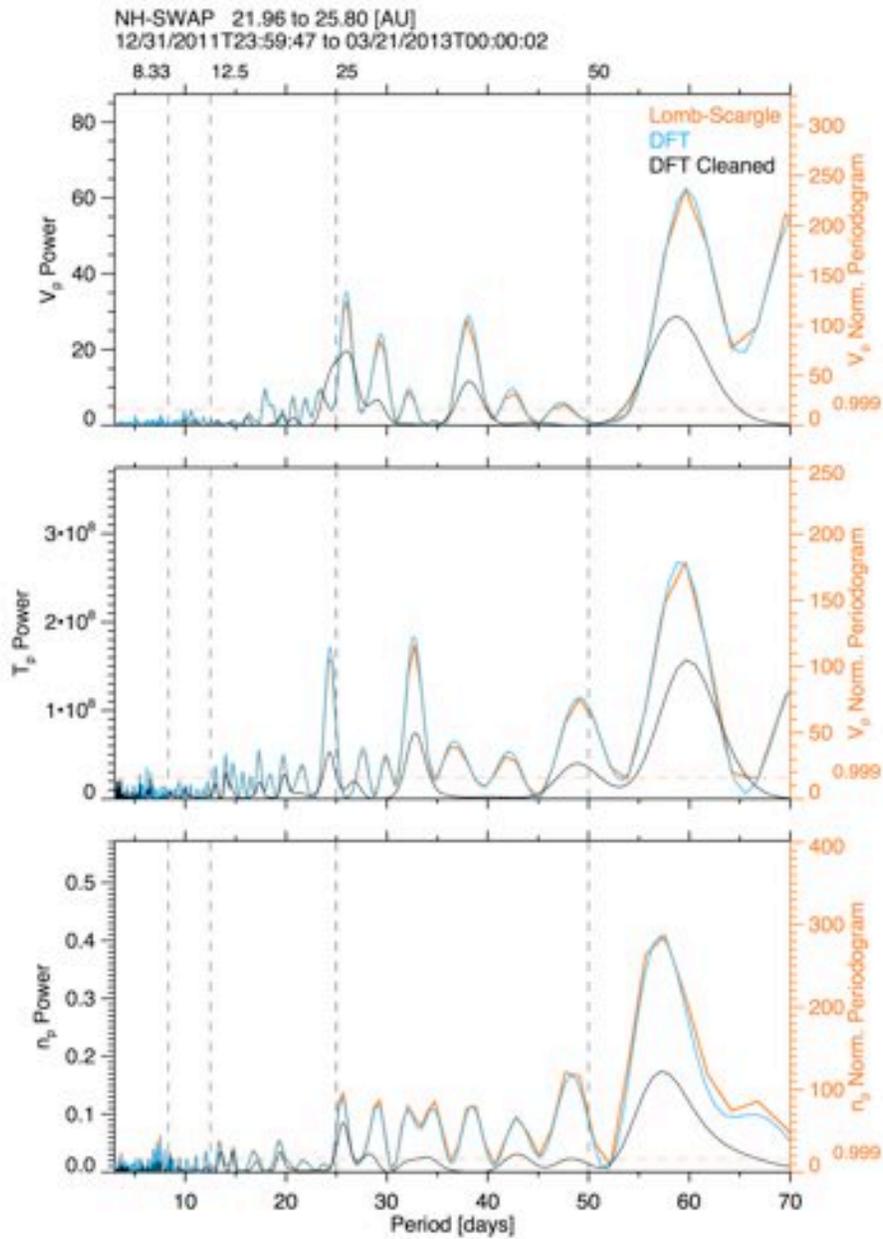

*Figure 25*: This figure has the same format as Figure 24, but here the periodograms are derived from SWAP data from 21.96 to 25.80 AU.

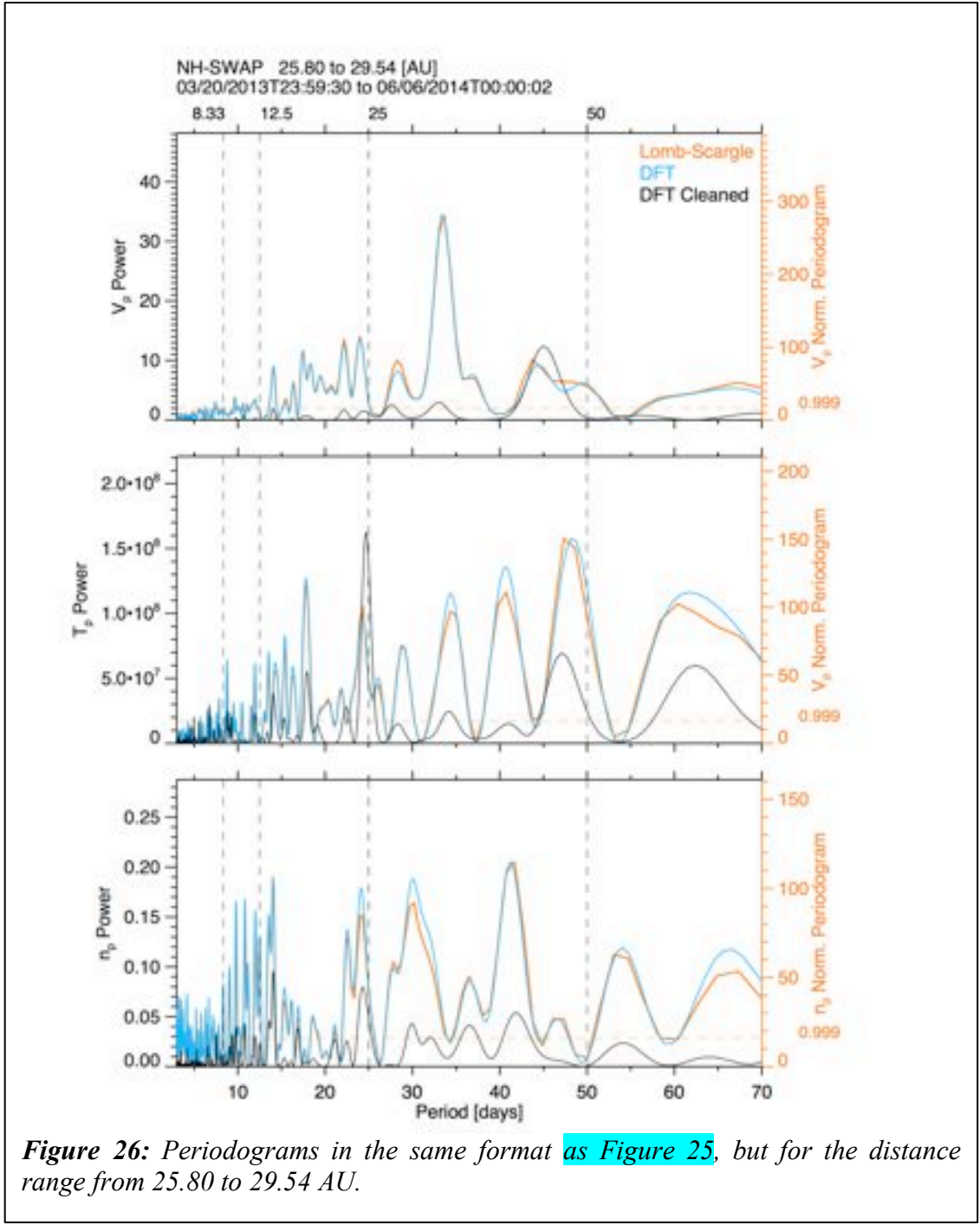

*Figure 26:* Periodograms in the same format as Figure 25, but for the distance range from 25.80 to 29.54 AU.

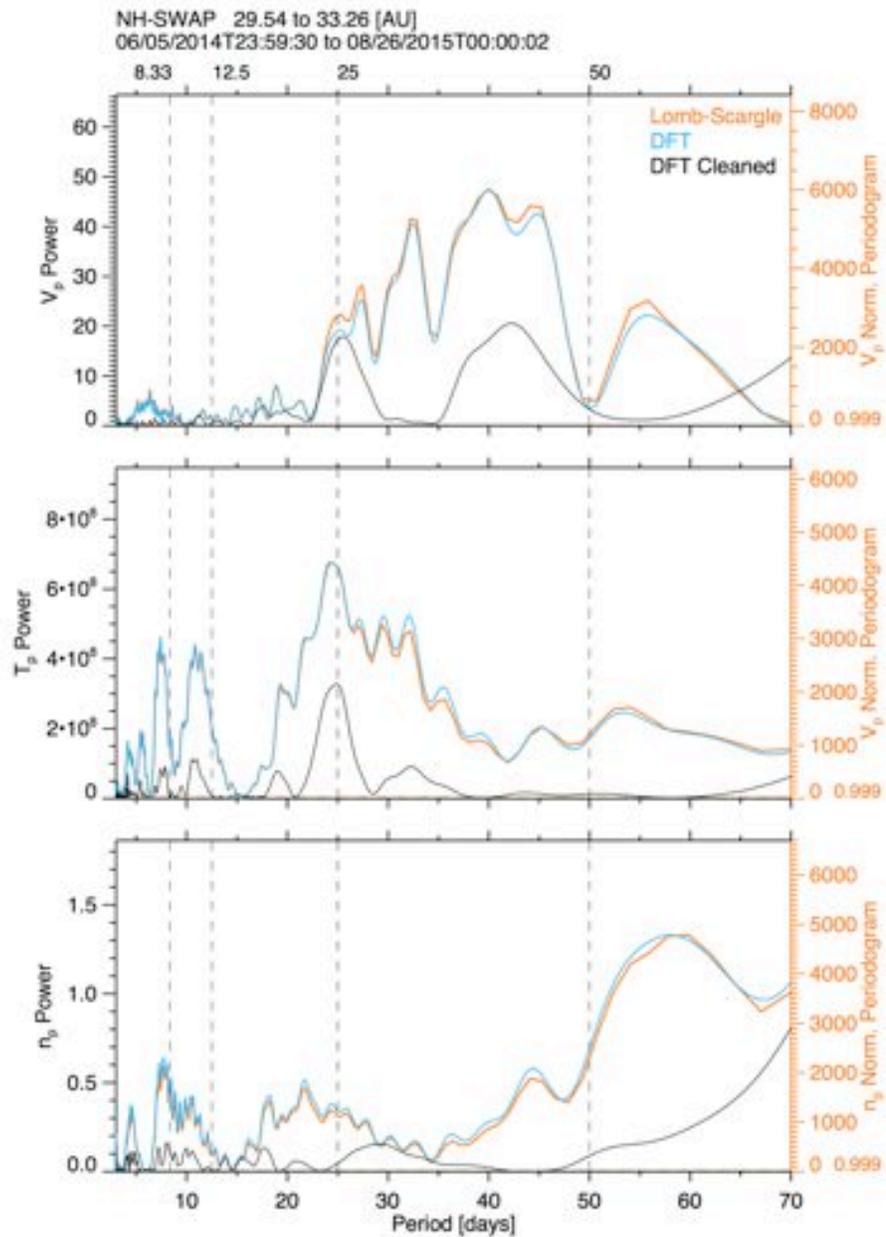

*Figure 27:* Periodograms in the same format as Figure 25, but for the distance range from 29.54 to 33.26 AU.

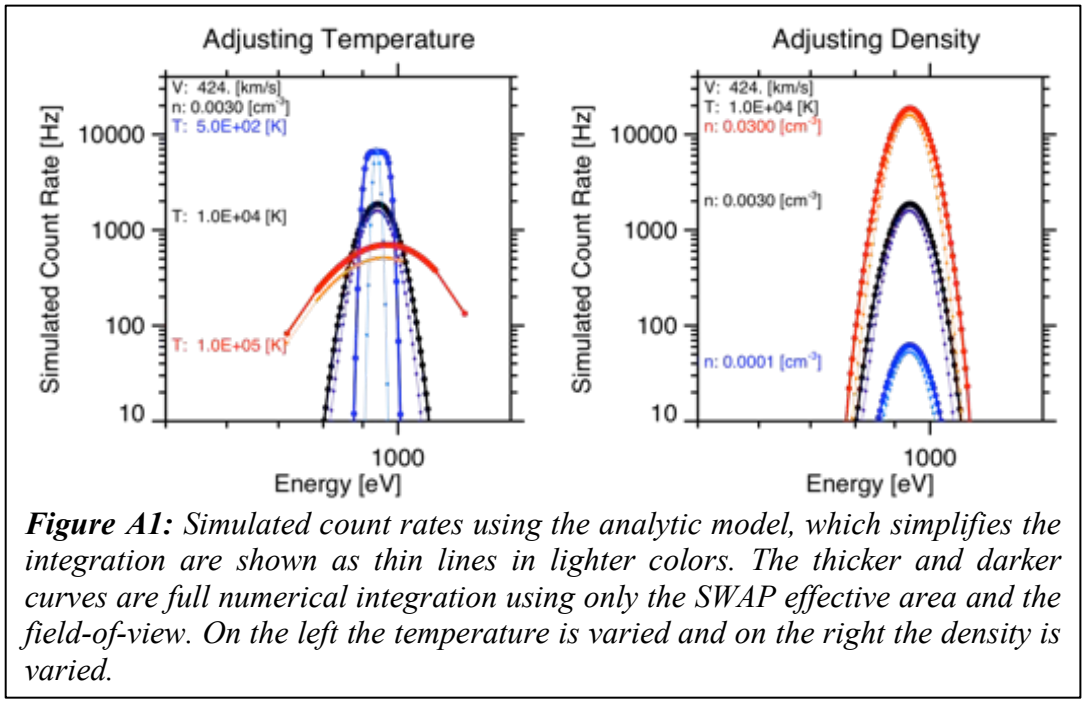

*Figure A1:* *Simulated count rates using the analytic model, which simplifies the integration are shown as thin lines in lighter colors. The thicker and darker curves are full numerical integration using only the SWAP effective area and the field-of-view. On the left the temperature is varied and on the right the density is varied.*

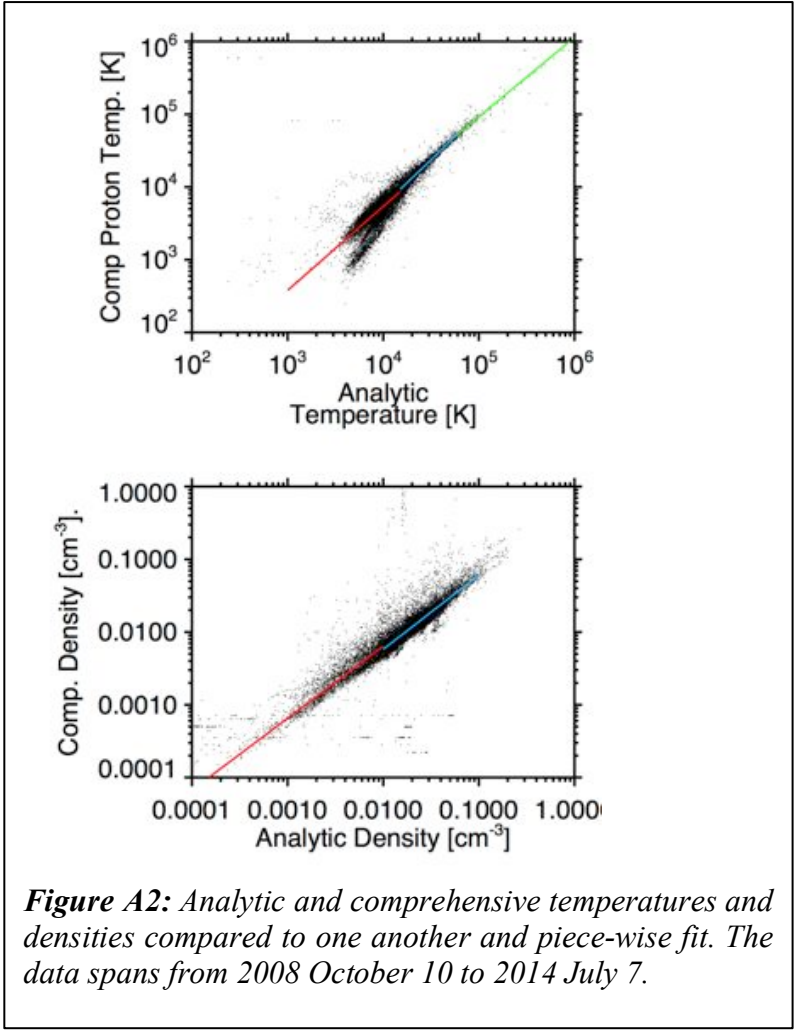

*Figure A2:* *Analytic and comprehensive temperatures and densities compared to one another and piece-wise fit. The data spans from 2008 October 10 to 2014 July 7.*